\documentclass[%
 reprint,
 amsmath,amssymb,
pra,
]{revtex4-2}
\usepackage[utf8]{inputenc}
\usepackage{graphicx}
\usepackage{dcolumn}
\usepackage{bm}
\usepackage{float}
\newcommand{\eq}[1]{Eq.\eqref{#1}}
\newcommand{\figref}[1]{Fig.~\ref{#1}}

\usepackage{xcolor}
\usepackage[maxfloats=256]{morefloats}
\DeclareUnicodeCharacter{0420}{~}
\DeclareUnicodeCharacter{0438}{~}     
\DeclareUnicodeCharacter{043E}{~}
\DeclareUnicodeCharacter{043B}{~}
\DeclareUnicodeCharacter{041A}{~}
\DeclareUnicodeCharacter{0430}{~}
\DeclareUnicodeCharacter{0441}{~}
\DeclareUnicodeCharacter{043A}{~}
\DeclareUnicodeCharacter{043F}{~}
\begin{document}

\preprint{APS/123-QED}
\title{A Theoretical Framework for Electromechanically Reinforced Brillouin Scattering in Integrated Photonic Waveguides  }

\author{ Ali Dorostkar}
 \affiliation{%
 $^{1}$ Fractal Group, Isfahan, Iran}%
 \email{m110alidorostkar@gmail.com}
\author{Sayyed Reza Mirnaziry }%

\affiliation{%
 $^{2}$  Electrical Engineering Department, University of Qom, Qom, Iran
}%


\begin{abstract}
 Current theoretical demonstration of the Stimulated Brillouin scattering (SBS) in  waveguides composed of centrosymmetric materials does not capture physics of the phenomenon in waveguides composed of non-centrosymmetric materials. The SBS in the latter problem, entails mutual coupling of the electric and acoustic waves due to piezoelectricity, and ends up to a different power conversion equation than that in a centrosymmetric material. In this study, we theoretically investigate Brillouin scattering in chip-scale waveguides with  non-centrosymmetric material when excited by an Inter-Digital-Transducer. We explain how Brillouin scattering is formulated in the presence of externally injected acoustic waves. Also we demonstrate the effect of this external signal on reinforcing Stimulated Brillouin scattering. As a case study we study SBS in a gallium arsenide nonowire. It is shown that the Stokes amplification due to electromechanically  reinforced SBS in this waveguide can grow several orders of magnitude higher than the values reported for the SBS in a silicon nanowire; This enables reducing the waveguide length required for a given Stokes amplification from centimeters to few hundred micrometers.  \\
\end{abstract}

\maketitle


\section{INTRODUCTION} 
Stimulated Brillouin Scattering (SBS) is a nonlinear process resulting from coherent interactions between photons and excited phonons in an optical propagation medium \cite{boyd2020nonlinear,Brillouin1922diffusion,mandelstam1926light,eggleton2019Brillouin,garmire2017perspectives}. There have been prominent developments in the last decade to tailor Brillouin interactions in chip-scale platforms; To enhance the SBS nonlinearities, various configurations have been studied among them nonowires - either suspended or on substrate- , slot and rib waveguides and photonic crystal fibers have shown promising potential to increase the SBS gain \cite{rakich2012giant, wolff2015stimulated,shin2013tailorable,poulton2013acoustic,van2015interaction,mirnaziry2016stimulated,gyger2020observation} for the purpose of various functionalities such as in  sensor, microwave photonics and lasing \cite{eggleton2019Brillouin}.
 \\ In harnessing SBS for integrated photonics the main challenge toward reaching higher values of  SBS gain is to effectively confine both optic and acoustic waves simultaneously in a waveguide cross-section. This underlines the importance of both waveguide materials and geometry to enable co-propagation of optical and acoustic waves \cite{pant2011chip}. These factors also influence the acoustic wave velocity and propagation length, hence change the Brillouin susceptibility \cite{poulton2013acoustic,otterstrom2023modulation}. 
 \noindent For instance, the optical modes in silicon nanowire on silica substrate is confined due to total internal reflection, however; due to the higher stiffness of silicon rather than $\rm SiO_2$ there is a leakaging of acoustic wave to the substrate. Chalcogenide glasses with relatively  high refractive index and low stiffness, are also promising candidates to realize total internal reflection for both optical and acoustics waves \cite{poulton2013acoustic}. Suspending the waveguide structures, despite the difficulty in their realization were also proposed to reach a mechanical isolation \cite{shin2013tailorable}. 
 Despite these successes, enabling confinement of photon and phonon in silicon or popular semiconductor platform with a feasible and robust manufacture still is  challenging. 
As the Brillouin optomechanical coupling is restricted by optical and mechanical dissipations. \\
\noindent Recently, a great attention is drawn toward achieving Brillouin scattering in chip scale geometries using electromechanical excitation of phonon  by  interdigital transducers (IDT) \cite{liu2019electromechanical,kittlaus2021electrically,mayor2021gigahertz}. A radio frequency (RF)-fed-IDT generates an external acoustic wave and inject it to SBS interaction medium. Provided that the applied signal is phase matched with the interested acoustic eigenfrequency of the waveguide, the Brillouin susceptability can be engineered by orders of magnitude \cite{otterstrom2023modulation}. Regarding this, the  electromechanical Brillouin scattering (EBS) is a process wherein a radio frequency  signal of generates acoustic waves, which is employed to contribute in power conversion of the injected optical pump \cite{liu2019electromechanical}. It is shown  that an acoustic wave at 16 GHz excited by  IDT  in an  aluminum nitride (AlN)  suspended  waveguide  scatters pump wave, which red-shifts it to  the anti-Stokes sideband. However, there is a discrepancy between theoretical results and those obtained by experiment on demonstrating the relevant Brillouin interactions \cite{liu2019electromechanical}.
\\
Current studies on SBS has also shown great interests on deploying acousto-optic platforms to harness piezo-optomechanical interactions. Piezolectric semiconductors have exhibited strong acoustoelectric interactions which has made them attractive to be used as IDTs \cite{kittlaus2021electrically,shao2019microwave,dahmani2020piezoelectric}. In\cite{kittlaus2021electrically} for instance, the acousto-optic modulation within silicon is demonstrated using IDT on the piezoelectric material AlN. Lithium niobate is another promising material for this platform which has shown a high mechanical Q factor and an efficient piezoelectric transduction \cite{shao2019microwave,dahmani2020piezoelectric}. 
In addition, several approaches are demonstrated to determine optimal condition for overlapping the electromechanically excited  phonon with a phonon generated  optomechanically in acousto-optic structures  \cite{balram2017acousto,jiang2020efficient,wu2020microwave}.
\noindent However, a general theoretical model is still needed to consider both EBS and SBS interactions in piezoelectric structures.
\\
Here, we provide a theoretical framework to study the effect of external piezoelectric forces induced by IDT on power conversion through the SBS in selected integrated photonic structures. We show how coupled mode equations for the pump and Stokes powers are modified in the presence of electromechanically injected phonons, and how SBS interactions form a distinguished power conversion between pump and Stokes. To demonstrate what we call Electromechanically Reinforced Stimulated Brillouin Scattering (ERSBS), we begin with the SBS formulation presented in \cite{wolff2015stimulated} and extend it to include the effect of piezoelectric forces. As we will show, piezoelectric force affects the SBS  gain and leads to a modification on the whole coupled mode equations. Also, the strength of piezoelectric force that is a function of piezoelectric power, is a critical parameter in  maximizing the Stokes amplification.

\section{PRELIMINARIES and definitions}
\begin{figure}
	\centering
	\includegraphics[width=1\linewidth]{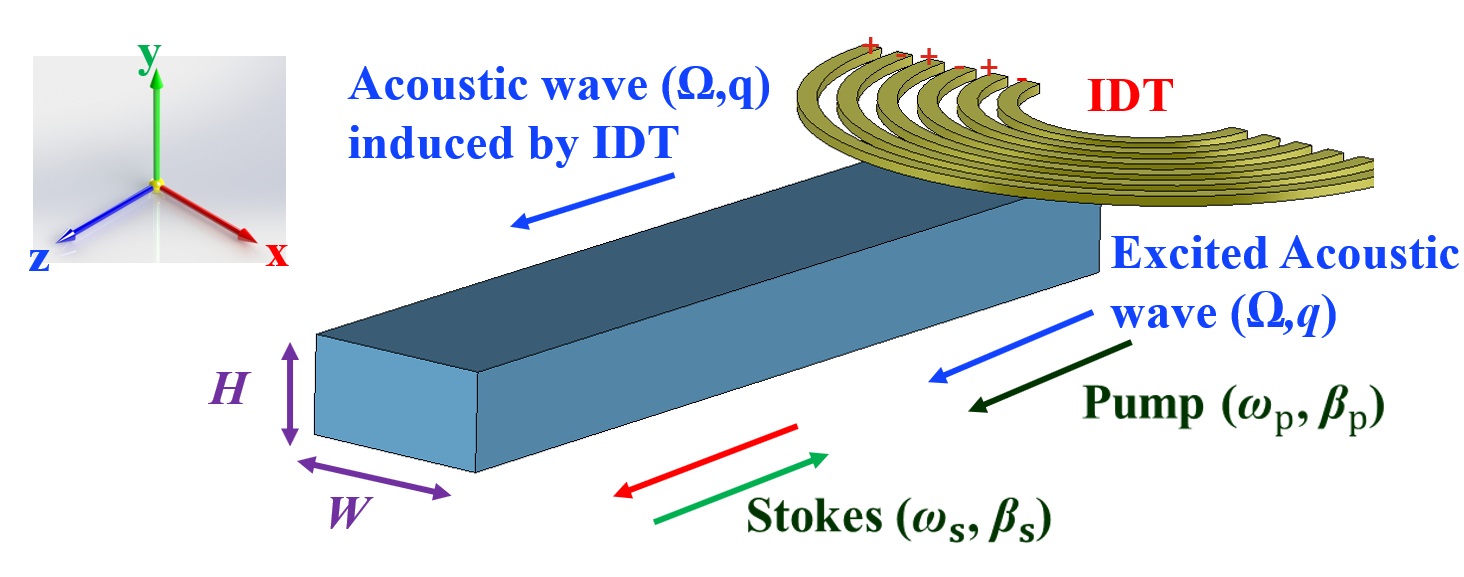}
	\caption{Schematic of the ERSBS interactions in a waveguide aligned along the z-axis with a cross section of $W$ (width) and $H$ (height). Backward ERSBS (Stokes wave represented by the green arrow) and forward ERSBS (Stokes wave in red) where pump and acoustic waves are represented by black and blue arrows, respectively.
	}
	\label{ESBS}
\end{figure}

We begin by introducing the geometrical and parameters of the interacting waves in establishing the coupled mode equations. Figure \ref{ESBS} shows a schematic of a nanowire composed of a piezoelectric material of rectangular cross section with a  width of $W$ and height of $H$ in the XY plane, capable of guiding light in at least one optical mode that is translationally invariant along the z-axis. As shown in the figure, an IDT is placed on the waveguide to induce acoustic waves. Assuming that pump and Stokes co-propagate along the externally induced acoustic wave, we call the process forward ERSBS. If instead pump and Stokes contra-propagate, while pump is in the direction of the external acoustic wave, a backward ERSBS occurs. 
The acoustic wave induced by IDT for forward or backward process has the same propagation direction with pump -or with the acoustic wave excited via pump-Stokes interactions.
An optical mode propagating through the nanowire at frequency of $\omega_i$ with a propagation constant $k_i$ (i= p (pump) or s (Stokes) ) can be expressed by 
\begin{align}
\mathbf E_i=A_i(z,t)\mathbf{\widetilde{e_i}(x,y)} e^{i(k_iz-\omega_i t)}+{\it c.c.},
\label{E_wave}
\end{align}
\noindent in which $A_i(z,t)$  is the envelope that varies slowly along the waveguide and $\mathbf{\widetilde{e_i}(x,y)}$ denotes the mode profile. The assumption of slowly envelopes in the coupled mode equations helps in simplifying them by neglecting of the higher order spacial derivative and dispersion terms. Note that the term $ c.c. $ in  \eq{E_wave} refers to complex conjugate components. Similarly, the acoustic displacement wave  at frequency of $\Omega$ with a propagation constant $q$ takes the following ansatz
\begin{align}
\label{U-centro}
\mathbf U= b(z,t) \mathbf{ \widetilde u (x,y)} e^{i(qz-\Omega t)}+{\it c.c.},
\end{align}
\noindent where $b(z,t)$ and $\mathbf{ \widetilde u (x,y)}$ are the acoustic envelope and the mode profile, respectively.  

\section {Coupled-Mode Equations in the presence of Centrosymmetric Materials} 

We first consider that the waveguide shown in Fig.\ref{ESBS} is composed of a centrosymmetric materials such as silicon. Then, the coupled-mode equations between pump, Stokes and the acoustic envelops are expressed by \cite{wolff2015stimulated}

\begin{eqnarray}
&\frac{\partial{A_{\rm p}}}{\partial{z}}=-\frac{i\omega_{\rm p} }{\mathcal P_{\rm p}}\widetilde{Q}_{\rm p}A_{\rm s}b,
\label{p}
\\
&\frac{\partial{A_{\rm s}}}{\partial{z}}=-\frac{i\omega_{\rm s} }{\mathcal P_{\rm s}}\widetilde{Q}_{\rm s}A_{\rm p}b^{*}\label{s},\\
&\frac{\partial b}{\partial z}+\alpha_{\rm ac} b=\frac{-i\Omega}{\mathcal P_{\rm ac}}\widetilde{Q}_{\rm b}A_{\rm p}A^{*}_{\rm s},
\label{acoustic_centro}
\end{eqnarray}
\noindent where $Q_i$ is an overlap integral, $ \omega_{\rm i}$ is the optical frequency of pump and Stokes, $ \Omega $ is the acoustic frequency and $\mathcal P_k$ $(\rm k = p , s,ac) $ is Poynting power for the optical and acoustic modes \cite{wolff2015stimulated}. The coefficient of $\alpha_{\rm ac}$ is the acoustic decay parameter.  The  overlap integral $Q_i$ is defined as 

\begin{equation}
\begin{split}
\widetilde{Q}_{\rm p}^{*}=\widetilde{Q}_{\rm s}=\widetilde{Q}_{\rm b}=\int (\mathbf{\widetilde u}^{*} \cdot\mathbf{\widetilde f }_{\rm EM})ds,
	\end{split}
	\end{equation}
\noindent where $(\mathbf{\widetilde f }_{\rm EM})$ denotes the  optical forces  \cite{wolff2015stimulated}.

\section {Coupled-Mode Equations of SBS in a Non-centrosymmetric Material} 
Now, we assume that the waveguide in Fig.\ref{ESBS} is composed of a non-centrosymmetric material.  We also limit our study to non-centrosymmetric materials with  piezoelectric properties; this means that crystals with  non-centrosymmetric cubic class 432 that do not possess piezoelectric properties are not considered here.  The piezoelectricity can influence acoustic parameters of the excited modes in the waveguide. Provided that an IDT is fed by a RF signal with an operating frequency equal to the interested acoustic mode, an extra phased-matched mechanical power is added to the one excited via SBS interactions. In the following, We first look at the relation between the piezoelectric signal and its resultant acoustic envelope, then use it to modify the coupled mode equations between pump and Stokes.
\subsection{Wave interactions in the presence of an externally injected piezoelectric signal }
We assume that the IDT generates an acoustic signal to a piezoelectric waveguide. We will add this piezoelectric signal to the wave equations because the acoustic mode must satisfy the required SBS phase matching condition. The induced mechanical vibrations in a piezoelectric waveguide leads an electromagnetic (EM) wave propagating along the waveguide at an identical frequency  to the acoustic signal (i.e. in the RF range) and vice versa. The electrical displacement wave ($\mathbf D_{\rm RF}$) corresponding to this wave is related to the mechanical strain by \cite{auld1973acoustic}
\begin{equation}
\begin{split}
\mathbf D_{\rm RF}={\bm{\bm \epsilon}}\cdot  \mathbf{E_{\rm RF}}+\mathbf e: \mathbf S+{\it c.c.},
\label{displacement_Strain}
\end{split}
\end{equation}
 \noindent where ${\bm \epsilon} $ is the permittivity, $\mathbf{E_{\rm RF}}$ is the electric wave, $\mathbf e$ is the piezoelectric stress tensor and $\mathbf S$ denotes strain produced by IDT and pump-Stokes interactions. In addition, the double dot indicates the product of a third rank tensor (i.e. the piezoelectric stress tensor) and a second rank tensor (i.e. the strain). It must to be mentioned that the extra term ($\mathbf e: \mathbf S$) comes from the material polarization. As strain is applied to the piezoelectric medium, this will change the orientation of dipole
arrangements in the material and leads to a net dipole moment per unit volume \cite{shepelin2021interfacial}. In addition, another RF signal can be generated from the pump and Stokes interference through second order nonlinearity, however; as shown in the appendix B its contribution is small enough to be neglected. 
  Now, the electromagnetic (EM) wave equation for a piezoelectric medium  in the RF domain can be described as: 
\begin{equation}
\begin{split}
\nabla\times\nabla\times\mathbf{E_{\rm RF}}=-\mu_0\frac{\partial^{2}\mathbf{D_{\rm RF}}}{\partial{t^{2}}}
\end{split}\label{Wave_eq_RF1}
\end{equation}\\
\noindent We note that in \eq{Wave_eq_RF1}, it is assumed that $ \mathbf{E_{\rm RF}} $ is induced by the IDT and impact by optical interactions through acoustic wave; in the presence of pump and Stokes signal, an extra phased matched RF signal is expected to be added to that generated by IDT. 
\\
Generally, the electric wave $ \mathbf{E_{\rm RF}} $ can be separated into a rotational and an irrotational part; The irrotational wave is called a quasi-static wave and can be expressed as a gradient of an scalar potential ($\Phi_{\rm RF}$) \cite{auld1973acoustic}. Note that with neglecting the rotational wave by taking the divergence as shown in \eq{11}  the quasi-static model is an appropriate
model for many piezoelectric problems [20].
\begin{equation}
\begin{split}
\mathbf{E_{\rm RF}}=\mathbf{E^{\rm rot.}_{\rm RF}}+\mathbf{E^{\rm irrot.}_{\rm RF}}=\mathbf{E^{\rm rot.}_{\rm RF}}-\nabla\Phi_{\rm RF}.
\end{split}
\end{equation}
\noindent The piezopotential signal ($\Phi_{\rm RF}$) is a time-varying scalar function, which varies along the waveguide with respect to $ \mathbf{E_{\rm RF}} $. Then, $\Phi_{\rm RF}$ can be expressed by
\begin{equation}
\begin{split}
\Phi_{\rm RF}(x,y,z,t)&=\Psi(z,t){\phi}(x,y)e^{i(qz-\Omega t)}+{\it c.c.}\\&=\Psi\widetilde{\phi}+{\it c.c.},
\label{Psi}
\end{split}
\end{equation}
\noindent where $\Psi$ is the potential envelope and $\widetilde{\phi}_{\rm RF}$ is the potential scalar mode. 
By taking the divergence over \eq{Wave_eq_RF1} and considering $\nabla \cdot (\bm{\bm \epsilon} \cdot \mathbf{E^{\rm rot.}_{\rm RF}})=0$ and $\mathbf{E^{\rm irrot.}_{\rm RF}}=-\nabla\Phi_{\rm RF}$ and after a rearrangement, we have
\begin{align}
\nabla \cdot\big(-\bm{\bm \epsilon}\cdot\nabla(\Psi\widetilde{\phi})+\mathbf e :\nabla (b\mathbf{\widetilde{u}})\big)=0.
\label{11}
\end{align}
\noindent  Now, by projecting $\widetilde{\phi}^*\times(-i\Omega)$ on \eq{11} and assuming  small signal approximation  together with neglecting higher order derivatives of $\Psi$, the relation between the piezopotential envelope and its corresponding acoustic envelope takes the following form\\
\begin{equation}
	\zeta_{\rm 0}\frac{\partial\Psi}{\partial z}+\alpha_{\rm ElM}\Psi=\frac{\partial b}{\partial z}+ \alpha_{\rm MEl}b,
\label{both_Psi-acoustic_text}
\end{equation}\\
{where  the coefficients $	\zeta_{\rm 0}$, $\alpha_{\rm ElM}$ and $\alpha_{\rm MEl}$ are  $\frac{M_{\rm RF}^{(1)}}{M_{\rm RF}^{(3)}}$, $\frac{M_{\rm RF}^{(2)}}{M_{\rm RF}^{(3)}}$ and $\frac{M_{\rm RF}^{(4)}}{M_{\rm RF}^{(3)}}$, respectively and ${M_{\rm RF}^{(i)}}$ (i =1,2,3,4) is an integral expressed by
  }
\noindent \\
\begin{equation}
\begin{aligned}
& {M_{\rm RF}^{(1)}}=(-i\Omega)\int \widetilde{\phi}^{*}\big(([\hat{a}_z]_{1\times3}\cdot (\bm{\bm \epsilon}\cdot\nabla\widetilde{\phi})+\nabla\cdot(\bm{\bm \epsilon}\cdot [\widetilde{\phi}]_{\rm z})\big)ds,\\
& {M_{\rm RF}^{(2)}}=(-i\Omega)\int \widetilde{\phi}^{*}\big(\nabla\cdot(\bm{\bm \epsilon}\cdot\nabla \widetilde{\phi})\big))ds,\\
& {M_{\rm RF}^{(3)}}=(-i\Omega)\int \widetilde{\phi}^{*}\big(([\hat{a}_z]_{1\times3}\cdot (\mathbf  e
:\nabla\mathbf{\widetilde{u}}))+\nabla\cdot(\mathbf e: [\mathbf{\widetilde{u}}])\big) ds,\\
&{M_{\rm RF}^{(4)}}=(-i\Omega)\int \widetilde{\phi}^{*}\big(\nabla\cdot(\mathbf e:\nabla \mathbf{\widetilde{u}})\big)ds,
\end{aligned}
\end{equation}
\noindent where $[\hat{a}_z]_{1\times3}$ is the unit vector $[0 , 0, 1]$, $[\widetilde{\phi}]_{\rm z}$ is defined as $[0 , 0, \widetilde{\phi}]$  and $[\mathbf{\widetilde{u}}]$  denotes $[0 ,0,\mathbf{\widetilde{u}_z},\mathbf{\widetilde{u}_y},\mathbf{\widetilde{u}_x}, 0]^{\rm T}$. Detail of derivations is provided in the appendix A.

{The ${M_{\rm RF}^{(i)}}$  coefficients are basically power terms in the piezoelectric medium; ${M_{\rm RF}^{(1)}}$  and ${M_{\rm RF}^{(2)}}$   are related to  the  electrical power and power per length, respectively, while 
${M_{\rm RF}^{(3)}}$   and ${M_{\rm RF}^{(4)}}$ describe  the  mechanoelectrical coupling power and power per length, respectively. The mechanoelectrical (electromechanical) coupling explains the amount of mechanical (electrical) energy that is converted to electrical (mechanical) energy. The energy conversion - from mechanical to electrical and vice versa -  can be interpreted by a decay factor as denoted with $\alpha_{\rm ElM}$ ($\alpha_{\rm MEl}$) in \eq{both_Psi-acoustic_text}. These energy conversions are also expressed by an electromechanical (mechanoelectrical) coupling coefficient $\kappa$ in \cite{auld1973acoustic}.
Generally, We can recognize four types of losses in a piezoelectric material during Brillouin interactions; they are corresponding to the imaginary parts of viscosity and piezoelectric and permittivity coefficients.   These power dissipation sources introduce an acoustic decay factor ($\alpha_{\rm ac}$),  coupling loss factors $\alpha_{\rm ElM}$  and $\alpha_{\rm MEl}$ - from electromechanical and mechanoelectrical processes, and a dielectric loss factor because of the material polarization  and conduction loss in the coupled mode equations.
Where only acoustics loss from viscosity is considered in this study and we neglect the others such as electromechanical, mechanoelectrical, dielectric and conduction losses \cite{gonzalez2016revisiting,yuan2011loss}. 
}

\subsection { Acoustic wave in a piezoelectric Material} 
We here derive an envelope acoustic differential equation from the acoustic wave equation  for a non-centrosymmetric medium in the presence of piezoelectric and optical forces.
The acoustic wave equation is expressed by \cite{auld1973acoustic}
\begin{align}
	\nabla\cdot \mathbf T+ \mathbf F_{\rm Piezo}+\mathbf F_{\rm EM}=\rho\partial^{2}_{\rm t}\mathbf U\label{conventional_elastic},
\end{align}
\noindent where $ \mathbf T$, $\mathbf F_{\rm piezo}$ and $\mathbf F_{\rm EM}$  are normal stress, piezoelectric force and the optically induced optical force, respectively. The total stress ($\mathbf T_{\rm tot}$) in piezoelectric material is divided to two parts of stress ($\mathbf T$) and piezoelectric stress ($\mathbf T_{\rm piezo}$)  and takes the following form \cite{auld1973acoustic}
\begin{equation}
\begin{split}
&\mathbf T_{\rm tot}=\mathbf T+\mathbf T_{\rm piezo}=\mathbf c:\mathbf S+ \bm\eta:\frac{\partial \mathbf S}{\partial t}-\mathbf e\cdot \mathbf E_{\rm RF},\\
&\mathbf T=\mathbf c:\mathbf S+ \bm\eta:\frac{\partial \mathbf S}{\partial t}, \\
&\mathbf T_{\rm piezo}=-\mathbf e\cdot \mathbf E_{\rm RF}, 
\label{T-centro}
\end{split}
\end{equation}
\noindent where $\mathbf c$ and $\bm\eta$ are stiffness and viscosity tensors, respectively.
The stress in a non-centrosymmetric material has an extra term in comparison to a centrosymmetric material. This term emerges as a piezoelectric force in the coupled mode equation
 \begin{widetext} 
\begin{equation}
\begin{split}
&\mathbf F_{\rm piezo}=\nabla \cdot\mathbf T_{\rm piezo} =\frac{\partial \Psi}{\partial z}\mathbf {\widetilde f}_{\rm 1,piezo}+\Psi\mathbf {\widetilde f}_{\rm 2,piezo}+{\it c.c.},
\label{F_piezo1}
\end{split}
\end{equation}
\end{widetext}
{where the $\mathbf{\widetilde f}_{\rm 1,piezo}$ and $\mathbf{\widetilde f}_{\rm 2,piezo}$  are the surface and volume piezoelectric forces, respectively.{It can be observed that in nano piezoelectric waveguides, impact of surface force play role in the overall force.}}
\noindent The piezoelectric overlap integral can be expressed by projecting the piezoelectric forces on the excited acoustic mode. 
 \begin{widetext} 
\begin{equation}
\begin{split}
&\widetilde{Q}_{\rm 1,piezo}=\int\mathbf {\widetilde u}^{*} \cdot\bigg(\big(([\hat{a}_z]_{3\times6}\cdot (\mathbf e\cdot\nabla\widetilde{\phi}))+\nabla\cdot(\mathbf e\cdot [\widetilde{\phi}]_{\rm z})\big)\bigg) ds=\int  (\mathbf {\widetilde u}^{*} \cdot \mathbf {\widetilde f}_{\rm 1,piezo}) ds,\\
&\widetilde{Q}_{\rm 2,piezo}=\int  \mathbf {\widetilde u}^{*} \cdot(\nabla\cdot(\mathbf e\cdot\nabla \widetilde{\phi}) ds =\int  (\mathbf {\widetilde u}^{*} \cdot \mathbf {\widetilde f}_{\rm 2,piezo}) ds.
\label{F_piezo2}
\end{split}
\end{equation}
\end{widetext}
where the matrix $[\hat{a}_z]_{3\times6}$ as shown in the appendix.
{\noindent The optical force $\mathbf F_{\rm EM}$ is predominantly originated from the electrostrictive (ES) force and the radiation pressure (RP) \cite{rakich2012giant}
\noindent and an appropriate ansatz for $\mathbf{F}_{\rm EM}$ with an optical force mode $\mathbf {\widetilde f}_{\rm EM}$ is \cite{ wolff2015stimulated} 
\begin{equation}
\mathbf F_{\rm EM}=\mathbf {F_{\rm ES}}+\mathbf {F_{\rm RP}}=A_{\rm p}A^{*}_{\rm s} \mathbf {\widetilde f}_{\rm EM}e^{-i\Omega t}+{\it c.c.},
\end{equation}}
\noindent By applying \eq {U-centro} and \eq {T-centro} in the acoustic wave equation and dropping higher order terms in the steady state regime and a projection on  the acoustic velocity mode $ \widetilde{\mathbf v}^{*} $ on \eq{differential_final-acoustic2} and then by knowing that the eigenmode 
$\mathbf{\widetilde{u}} $ (i.e.  $b=1$) satisfies the acoustic wave equation (more details in the appendix C), the differential equation governing the variations of the acoustic envelope is given by
\begin{widetext}
	\begin{equation}
	\begin{split}
	\frac{\partial b}{\partial z}+\alpha_{\rm ac} b=\frac{-i\Omega \widetilde{Q}_{\rm b}}{\mathcal  P_{\rm ac}+i \mathcal P_{\rm LOSS}}A_{\rm p}A^{*}_{\rm s}+\frac{-i\Omega \widetilde Q_{\rm 1,piezo}}{\mathcal  P_{\rm ac}+i \mathcal P_{\rm LOSS}}\frac{\partial \Psi}{\partial z}+\frac{-i\Omega \widetilde Q_{\rm 2,piezo}}{\mathcal  P_{\rm ac}+i \mathcal P_{\rm LOSS}}\Psi,
	\label{final_acoustic_eq22}
	\end{split}
	\end{equation}
	\end{widetext}
\noindent where $\alpha_{\rm ac}$, $\mathcal  P_{\rm ac}$, $\mathcal P_{\rm LOSS}$, $ \widetilde{Q}_{\rm i,piezo}$ (i=1,2) and $ \widetilde{Q}_{\rm b}$ are the linear loss coefficient, acoustic Poynting power, acoustic loss power and overlap integrals of piezoelectric and opto-acoustic, respectively. These coefficients  are calculated in the appendix.
With \eq{both_Psi-acoustic_text} and and rearrangement it based on the $\frac{\partial b}{\partial z}$  and then by substituting in \eq {final_acoustic_eq22}, the  acoustic envelope is 
\begin{equation}
	\begin{split}
&b=\gamma_1 A_{\rm p}A^{*}_{\rm s}+\gamma_2\frac{\partial \Psi}{\partial z}+\gamma_3\Psi,
\label{exact_bb}
\end{split}
	\end{equation}
  where $\gamma_i$ (i=1,2,3) are\\
 	\begin{equation}
	\begin{split}
&\gamma_1=\frac{-i\Omega \widetilde{Q}_{\rm b}}{(\alpha_{\rm ac}-\alpha_{\rm MEl}) \mathcal  P_{\rm ac}},\\& \gamma_2=(\frac{-i\Omega \widetilde Q_{\rm 1,piezo}}{(\alpha_{\rm ac}-\alpha_{\rm MEl})\mathcal  P_{\rm ac}}-\frac{\zeta_{\rm 0}}{(\alpha_{\rm ac}-\alpha_{\rm MEl})}),\\&\gamma_3=(\frac{-i\Omega \widetilde Q_{\rm 2,piezo}}{(\alpha_{\rm ac}-\alpha_{\rm MEl})\mathcal  P_{\rm ac}}-\frac{\alpha_{\rm ElM}}{(\alpha_{\rm ac}-\alpha_{\rm MEl})}),
\label{gamma_i_text1}
\end{split}
\end{equation}\\ 
{The acoustic envelope is a function of optical interaction between pump and Stokes, the strength of piezopotential signal and its variation through the direction of propagation. Note that pump and Stokes change the optical forces while the piezopotential term impact the piezoelectric force per volume and it spacial variation from surface piezoelectric force. Furthermore,  the mechanoelectrical coupling ($\alpha_{\rm MEl}$) acts as a loss for direct optical interaction of pump and Stokes. This means that, by generating an acoustic wave due to Brillouin interactions, a fraction of acoustic energy is transferred to the electrical domain (due to the mechanoelectrical effect). This introduces a loss term as shown in \eq {gamma_i_text1}. In the case of SBS in centrosymetric materials however,  the coefficients $\gamma_2$ and $\gamma_3$  and $\alpha_{\rm MEl}$ - in $\gamma_1$ -   become zero. }   
 \subsection { Optical Waves in a Non-centrosymmetric Material}
  In a waveguide in the present of both pump and Stokes, the total electric field can be expressed as a superposition of pump and stokes waves ($\mathbf E=\mathbf E_{\rm p}+\mathbf E_{\rm s}$).
In a perturbed waveguide, each field component component e.g. $\mathbf E_{\rm p}$ has two parts; a non-perturbed and a perturbed term due to the interaction with acoustic and Stokes waves \cite{wolff2015stimulated}. The latter, leads to temporal change in the refractive index. However, this is not the only source of change in refractive index of a piezolectric material, as here the electrooptic effect (here Pockels effect) can change the refractive index as well. Thus, the total refractive variation to distort the optical wave can be described as \cite{biaggio1999nonlocal}
\begin{equation}
\Delta(\frac {1}{\bm \epsilon})={\bm {r}}\cdot \mathbf E_{\rm RF}+ \mathbf p : \mathbf S\approx\mathbf p : \mathbf S,
\end{equation}
\noindent where $\bm {r}$ and $\mathbf p $ are electrooptic and photoelastic tensors, respectively. 
We should note that the RF signal generated through second order nonlinearity can interact with Stokes (pump) and thus create a pump (Stokes) wave ($A_{\rm p} \propto A_{\rm s}\Psi$). However with more details shown in the appendix B,
we neglect the impact of electrooptic effect and perturbation of optical wave by the RF signal ($\Psi\Delta\mathbf{\mathbf{\widetilde{e}_{\rm 2,i}}}$) and assume it is negligible, therefore, the whole of the RF power transformed into acoustic domain through the electromechanical effect. Then, total electric field based on the perturbation theory can be written as follows\\

\begin{equation}
\label{perturbation}
\begin{split}
\mathbf E=&A_{\rm p}\mathbf{\mathbf{\widetilde{e}_{\rm p}}}+A_{\rm s}b\Delta\mathbf{\mathbf{\widetilde{e}_{\rm 1,p}}}+A_{\rm s}\Psi\Delta\mathbf{\mathbf{\widetilde{e}_{\rm 2,p}}}+A_{\rm s}\mathbf{\mathbf{\widetilde{e_{\rm s}}}}+A_{\rm p}b^{*}\Delta\mathbf{\mathbf{\widetilde{e}_{\rm 1,s}}}\\&+A_{\rm p}\Psi^{*}\Delta\mathbf{\mathbf{\widetilde{e}_{\rm 2,s}}}+{\it c.c.}\\
=&A_{\rm p}\mathbf{\mathbf{\widetilde{e}_{\rm p}}}+A_{\rm s}b\Delta\mathbf{\mathbf{\widetilde{e}_{\rm 1,p}}}+A_{\rm s}\mathbf{\mathbf{\widetilde{e_{\rm s}}}}+A_{\rm p}b^{*}\Delta\mathbf{\mathbf{\widetilde{e}_{\rm 1,s}}}+{\it c.c.},
	\end{split}
\end{equation}
\noindent where $\Delta \mathbf{\widetilde{e_i}}$ is equal to $\Delta \mathbf{{e_i}}(x,y)e^{i(\beta_i z-\omega_i t)}$.
With this assumption, the formulation of governing envelope equations of pump and Stokes for non-centrosymmetric is the same with \eq{p} and \eq{s} with the differences in the results of $b$ and $\mathbf{\widetilde u}$. In non-centrosymmetric due to piezoelectric effect the acoustic mode profile and the envelope can be different from those in a centrosymmetric material. 
 \subsection { Coupled Mode Equations}
 \noindent Assuming equations of (\ref{p}), (\ref{s}), (\ref{both_Psi-acoustic_text}) and (\ref{final_acoustic_eq22}) the general coupled mode equations for optical, piezopotential and acoustic can be obtained. Depending on the presence and/or strength of acoustic and piezopotential envelopes, three different coupled mode interactions are conceivable. The corresponding envelope equations of each case is tabulated in Table 1.
\begin{widetext}
	\begin{center}
		\begin{tabular}{c | c| c 
		}
					\multicolumn{3} {  c  }{Table.1: The general coupled mode equations associated with different  Brillouin scattering scenarios}\\
		\\
			SBS \cite{wolff2015stimulated} & EBS& ERSBS\\
			\hline
			$\frac{\partial{A_{\rm p}}}{\partial{z}}=\frac{-i\omega_{\rm p} \widetilde{Q}_{\rm p}}{\mathcal P_{\rm p}}A_{\rm s}b$ &$\frac{\partial{A_{\rm p}}}{\partial{z}}=\frac{-i\omega_{\rm p} \widetilde{Q}_{\rm p}}{\mathcal P_{\rm p}}A_{\rm s}b$& $\frac{\partial{A_{\rm p}}}{\partial{z}}=\frac{-i\omega_{\rm p} \widetilde{Q}_{\rm p}}{\mathcal P_{\rm p}}A_{\rm s}b$\\
			$\frac{\partial{A_{\rm s}}}{\partial{z}}=\frac{-i\omega_{\rm s} \widetilde{Q}_{\rm s}}{\mathcal P_{\rm s}}A_{\rm p}b^{*} $ &$\frac{\partial{A_{\rm s}}}{\partial{z}}=\frac{-i\omega_{\rm s} \widetilde{Q}_{\rm s}}{\mathcal P_{\rm s}}A_{\rm p}b^{*} $&$\frac{\partial{A_{\rm s}}}{\partial{z}}=\frac{-i\omega_{\rm s} \widetilde{Q}_{\rm s}}{\mathcal P_{\rm s}}A_{\rm p}b^{*} $\\
			$\frac{\partial b}{\partial z}+\alpha_{\rm ac} b=\frac{-i\Omega\widetilde{Q}_{\rm b}}{ \mathcal P_{\rm ac}}A_{\rm p}A^{*}_{\rm s} $& $\frac{\partial b}{\partial z}+\alpha_{\rm ac} b=\frac{-i\Omega \widetilde Q_{\rm 1,piezo}}{\mathcal  P_{\rm ac}}\frac{\partial \Psi}{\partial z}+\frac{-i\Omega \widetilde Q_{\rm 2,piezo}}{\mathcal  P_{\rm ac}}\Psi $ & $\frac{\partial b}{\partial z}+\alpha_{\rm ac} b=\frac{-i\Omega \widetilde{Q}_{\rm b}}{\mathcal  P_{\rm ac}}A_{\rm p}A^{*}_{\rm s}+\frac{-i\Omega \widetilde Q_{\rm 1,piezo}}{\mathcal  P_{\rm ac}}\frac{\partial\Psi}{\partial z}$\\ &  &
			+ $\frac{-i\Omega \widetilde Q_{\rm 2,piezo}}{\mathcal  P_{\rm ac}}\Psi $\\
		 &$\frac{\partial b}{\partial z}+ \alpha_{\rm MEl}b=	\zeta_{\rm 0}\frac{\partial\Psi}{\partial z}+\alpha_{\rm ElM}\Psi $&$\frac{\partial b}{\partial z}+ \alpha_{\rm MEl}b=	\zeta_{\rm 0}\frac{\partial\Psi}{\partial z}+\alpha_{\rm ElM}\Psi$\\
			\end{tabular}
	\end{center}
\end{widetext}
 The first case is when there is no external injection of acoustic waves through the IDT. Thus, conventional SBS takes place. The second case refers to the situation that IDT induces power and its corresponding acoustic power is much stronger than that produced via pump-Stokes interactions. We call this, electromechanically induced Brillouin scattering (EBS). Finally the third case describes the situation that IDT is on and its induced acoustic signal is comparable with that generated via pump-Stokes interactions, a scenario that we call Electromechanically Reinforced SBS (ERSBS).
In deriving the following differential equation for the envelopes, we note that $\mathcal P_{\rm LOSS}$ is assumed to be zero ($\mathcal P_{\rm LOSS}$=0).

\subsection { Power Conversions in a piezoelectric waveguide}

{The governing equations expressing the variations of pump, Stokes and Piezoelectric powers are as follows (details of derivations can be found in the appendix E)}
 \begin{widetext}
	\begin{equation}
	\begin{split}
&\frac{\partial{P_{\rm p}}}{\partial{z}}=-\Upsilon_{\rm 0} P_{\rm p}P_{\rm s}-\Upsilon_{\rm 1}  P_{\rm piezo}{P_{\rm p}}+\Upsilon_{\rm 1} P_{\rm piezo}{P_{\rm s}}-\Upsilon_{\rm 2}  P_{\rm piezo}{P^{2}_{\rm p}}-\Upsilon_{\rm 2}  P_{\rm piezo}{P^{2}_{\rm s}}+2\Upsilon_{\rm 2}  P_{\rm piezo}{P_{\rm p}}{P_{\rm s}}-\Upsilon_{\rm 3}  P_{\rm piezo},\\
&\frac{\partial{P_{\rm s}}}{\partial{z}}=\Upsilon_{\rm 0} P_{\rm p}P_{\rm s}+\Upsilon_{\rm 1}  P_{\rm piezo}{P_{\rm p}}-\Upsilon_{\rm 1}  P_{\rm piezo}{P_{\rm s}}+\Upsilon_{\rm 2}   P_{\rm piezo}{P^{2}_{\rm p}}+\Upsilon_{\rm 2}  P_{\rm piezo}{P^{2}_{\rm s}}-2\Upsilon_{\rm 2}  P_{\rm piezo}{P_{\rm p}}{P_{\rm s}}+\Upsilon_{\rm 3}  P_{\rm piezo},\\
&\frac{\partial{P_{\rm piezo}}}{\partial{z}}=\Upsilon_{\rm piezo} P_{\rm piezo}( P_{\rm p}- P_{\rm s}),
\label{p_b_text}
	\end{split}
	\end{equation}
\end{widetext}
{\noindent where  $\Upsilon_i$  (i=0:3,piezo) are ERSBS (EBS) coefficients and calculated in the appendix.}
	

\section{Discussion}
\begin{figure}
\centering
\includegraphics[width=1\linewidth]{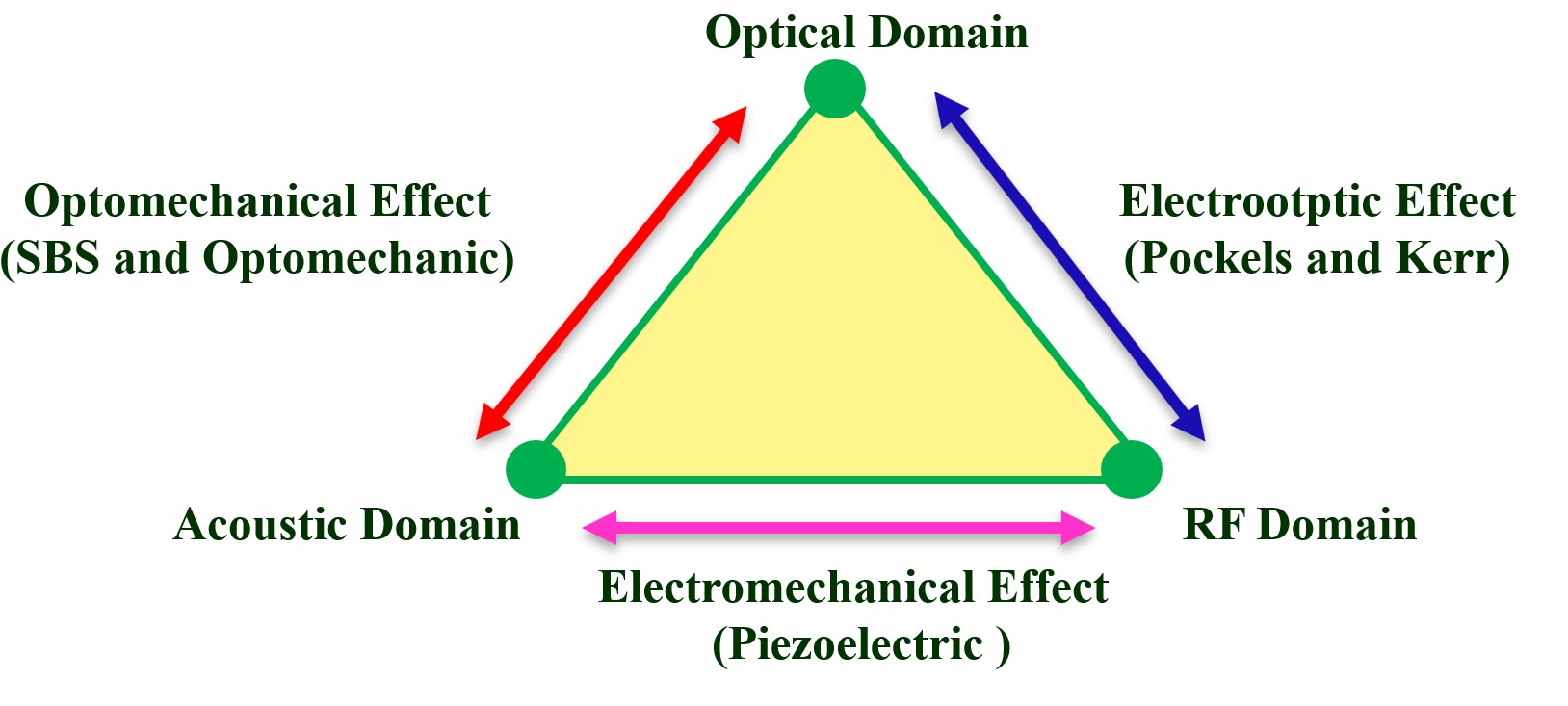}
	\caption{Conversion process of optical, acoustic and RF domains.
	}
	\label{tringle}
\end{figure}

\begin{figure}
	\centering
	\includegraphics[width=.8\linewidth]{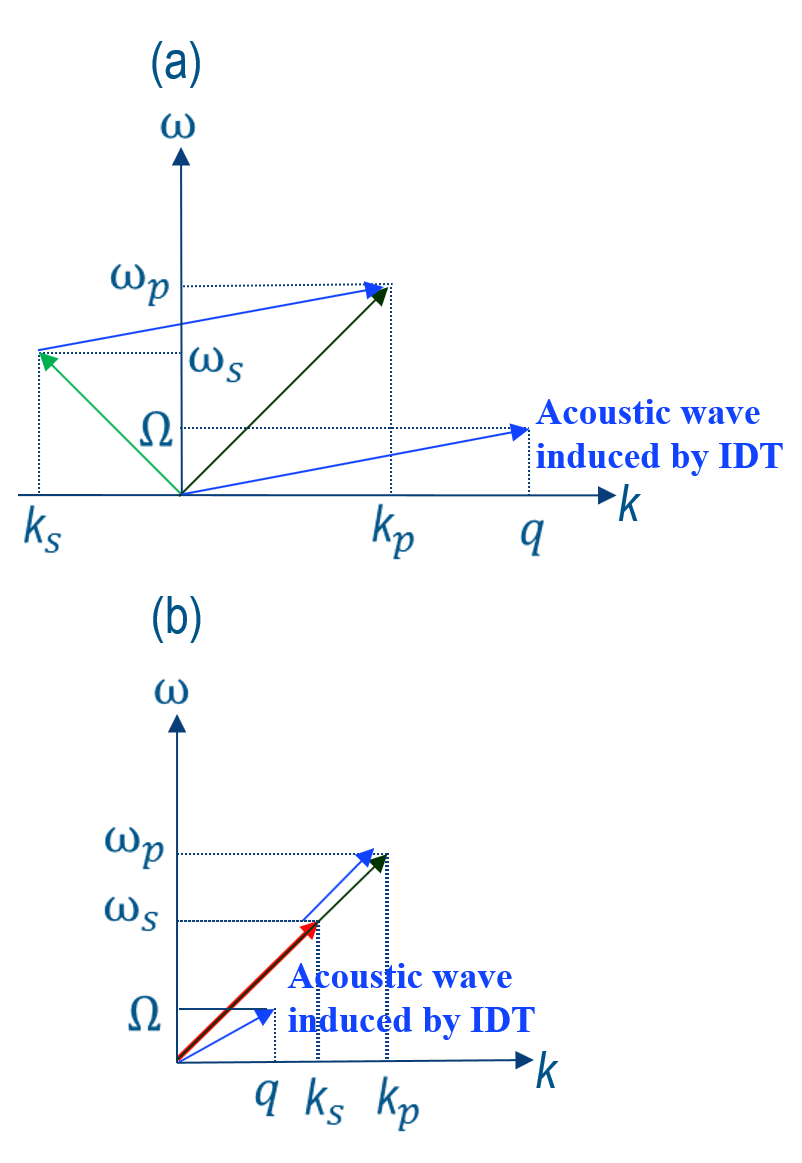}\\
		\caption{The dispersion diagram of the ERSBS (a) Backward (b) Forward process. Black and green (red) arrows indicate pump and backward (forward) Stokes, respectively and Blue arrows shows the acoustic dispersion diagram.
	}
	\label{dispersion_diagram}
\end{figure}

 The mutual effect of the RF signal - excited in a piezoelectric material by the IDT -  and the optical waves can be explained by a direct electrooptic effect, together with an indirect acoustic mediated process. As shown in \figref{tringle}, acoustic waves propagated in the structure are influenced by the strength of optical waves through SBS or other optomechanical phenomena such as piezo optomechanical cavity. The acoustic wave may also be excited  externally by IDT through piezoelectric effect separately. In our study, we showed how an acoustic wave excited by pump and Stokes waves through SBS effect and externally injection of phonon by IDT contribute in power conversion from pump to Stokes. We here neglect the electrooptic effect for the SBS process because the generated RF signal through $\chi^{(2)}$ effect is weak as shown in appendix B. Then, we move through the indirect interaction between the RF signal and the optical waves of \figref{tringle}, the two step process which we call electro-optomechanical interaction. The electrooptic side for our research is disconnected. 
Figure 3 shows the general SBS process for the intra mode coupling case. Intra mode is referred carrying of the identical mode of optical waves while inter mode is for different modes. Within the backward intra mode SBS (\figref{dispersion_diagram} (a)) the excited acoustic modes have large-wavenumbers and strong variations along the waveguide direction, which resemble them to a nearly longitudinal mode. In forward intra mode SBS (\figref{dispersion_diagram} (b)) the Stokes is co-propagates of optical and acoustics wave.  Therefore, due to conservation of momentum, acoustic modes occur near the acoustic propagation constant $q\approx0$ where the acoustic modes do not propagate along the waveguide axis any more (they possess the form of shear waves).
Thus, the modes have strong transverse components. Without considering the type of scattering, energy and momentum conservation laws make a constraint to prepare the phase-matching condition on the opto-acoustic interactions. The distinguishing phase-matching conditions result in different sets of coupled wave equations with unique dynamics. For reinforcing the SBS, an acoustic wave with same frequency ($\Omega$) and wave vector ($q$) as shown in \figref{dispersion_diagram} externally is induced to the waveguide by IDT, which enhances the excited acoustic wave due to pump-Stokes interactions.
{The physics behind the described electro-optomechanical interactions  can better be understood from a closed form expression for the acoustic envelope where more details about deriving equation are in the appendix G. In this regard, the acoustic envelope take the following expression 
 \begin{equation}
\begin{split}
b\approx\frac{-i\Omega Q_{\rm b}}{\alpha_{\rm ac}\mathcal P_{\rm ac}+i\Omega \frac{\alpha_{\rm MEl}}{\alpha_{\rm ElM}}\widetilde Q_{\rm 2,piezo}}  A_{\rm p}A^{*}_{\rm s},
\label{b_quantum111}
\end{split}
\end{equation}
As can be seen the impact of $\frac{\partial \Psi}{\partial z}$ is disappeared and the  acoustic envelope does not depend on $\widetilde Q_{\rm 1,piezo}$. However, the volume term contributes in the overall acoustic envelope when piezoelectricity exists in the waveguide medium. This leads to a simplified expression (in comparison with  the general formula shown in \eq{p_b_text}) for the SBS power conversion in piezoelectric waveguides. We note that \eq{b_quantum111} is closely similar to the one obtained for the acoustic envelope in a piezoelectric medium via quantum approach (see \cite{otterstrom2023modulation}) in which the dissipation rate in an electro-optomechanical structure takes a complex value. 
Electromechanical phonon gain directly modifies the nonlinear optical susceptibility through an enhancement in the effective Brillouin gain coefficient, which is inversely proportional to the phonon dissipation rate. These enhanced dynamics have important consequences for Brillouin-based devices such as amplifiers, lasers, nonreciprocal devices, optomechanical delay among others—especially in chip scale systems where the accessible levels of Brillouin gain have been historically limited \cite{otterstrom2023modulation}.}\\

\section{Case Study: ERSBS in Gallium Arsenide Nanowire}
\begin{figure}[H]
	\centering
	\includegraphics[width=1\linewidth]{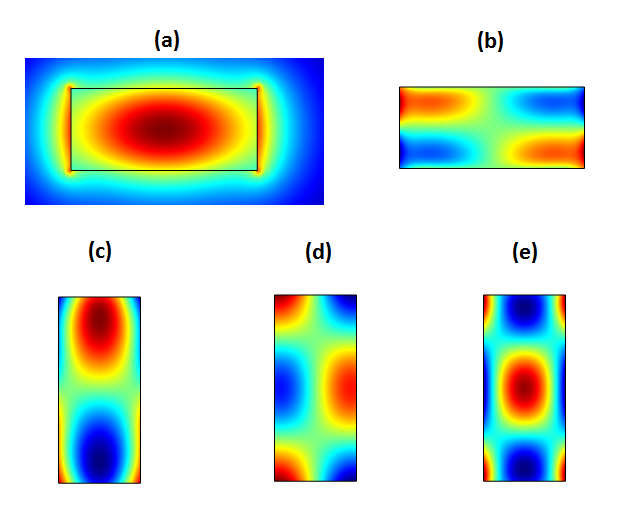}\\
		\caption{Mode profiles of norm of optical waves ($\mathbf{|E|})$ at 1550 nm (a) piezopotential mode ($\widetilde{\phi}$) at 66.05 [$\rm{GRad/s}]$ (b) acoustic mode in x direction ($\mathbf{\widetilde{u}_x}$) (c) acoustic mode in y direction ($\mathbf{\widetilde{u}_y}$)(d) acoustic mode in z direction ($\mathbf{\widetilde{u}_z}$) (e) for a GaAs waveguide with a cross section of 220 nm$\times$ 500 nm.
	}
	\label{modes_profile}
\end{figure}

\begin{figure}[H]
	\centering
	\includegraphics[width=1\linewidth]{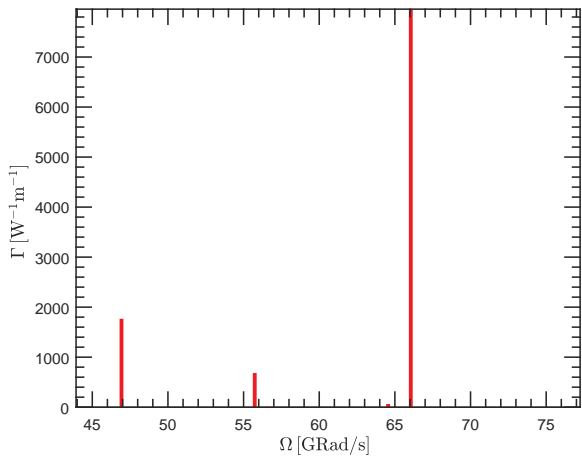}
			\caption{ Backward SBS gain without IDT.} 
			\label {SBS_gain}
	\end{figure}

We here solve the coupled mode equations in \eq{p_b_text} from numerical approaches. The applied method is different depend on the forward or to study the backward ERSBS process in a GaAs nonawire. In this study, we focus on  backward ERSBS where  pump and Stokes are counter propagated along the waveguide which leads to differential equations with two points boundary value problem.
A backward ERSBS in a GaAs nonowire with a cross section of 500 nm (width) and 220 nm (height), and refractive index of 3.37 at the wavelenght 1550 nm is demonstrated.
The fundamental optical mode and  acoustic and piezopotential modes at the corresponding frequency are depicted in \figref{modes_profile}. 
First, the SBS gain without IDT is investigated.
As shown in \figref{SBS_gain}, a maximum gain of 7957 $\rm {W^{-1}m^{-1}}$ occurs at $\Omega$= 66.05 [$\rm{GRad/s}]$. Next, the IDT becomes activated and launches a designed piezoelectric power to the structure. It must be mentioned that the acoustic loss coefficient ($\alpha_{\rm{ac}}=\frac{q}{Q_{\rm{m} }}$) is calculated by assuming a mechanical Q factor ($Q_{\rm{m}}$) of 1000 with $q= 1.51\times 10^{7}$ $\rm{Rad/m}$.

It can be observed from Table 2 that SBS gain from interaction of pump and Stokes has dramatically reduced due to presence of piezoelectric effect through $\Upsilon_0$ coefficient.
This impact is presented as loss for the proposed structure and reducing the parameter $\Upsilon_0$  significantly.  This ratio takes a complex value with large imaginary part for the proposed acoustic mode. The real part of this ratio is -6.033$\times$ 10$^{-8}$ $\rm{[m^{-1}]}$, while the imaginary part reaches -2.93$\times$ 10$^{7}$ $\rm{[m^{-1}]}$, which leads a significant reduction of the $\Upsilon_0$.

\begin{figure}[H]
	\centering
	\includegraphics[width=1\linewidth]{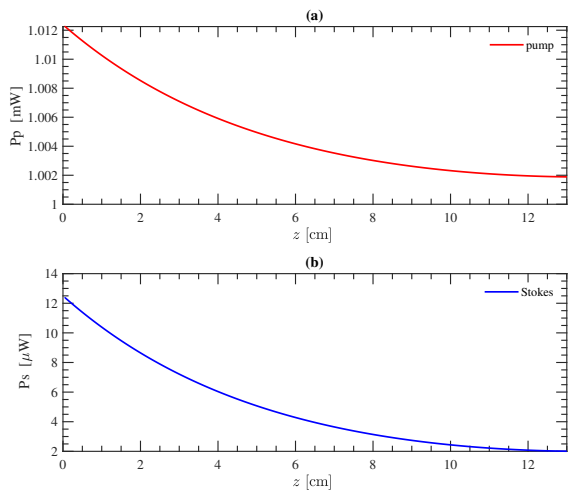}
			\caption{Power conversion  without IDT   with  initial power values of  (a) pump, 
 (b) Stokes and of 1.012 $\rm mW$ and 2 $\rm \mu W$  with a waveguide length of 13 cm.}
			\label{SBS13}
	\end{figure}

  \begin{figure}[H]
	\centering
	\includegraphics[width=1\linewidth]{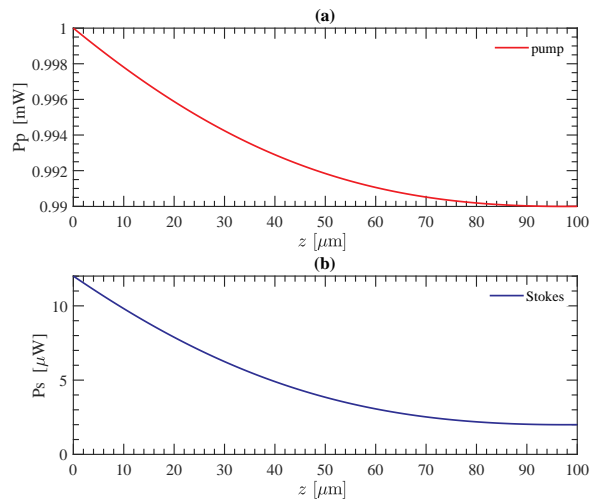}
	\caption{Power conversion with IDT with initial power values of (a) pump, (b) Stokes and piezoelectric of 1 $\rm mW$, 2 $\rm \mu W$ and 1 $\rm\mu W$ with a waveguide length of  100 $\rm\mu m$.}
	\label{ERSBS100} 
	\end{figure}
 
The power conversion for both scenarios ( i.e. SBS in the presence and absence of IDT) are shown in \figref{SBS13} and \figref{ERSBS100}.
It is assumed that the initial values for the pump and piezoelectric power  are about 1 $\rm mW$ and 1 $\rm\mu W$ at the beginning of the waveguide respectively while the 
\noindent initial value of Stokes power at the end of waveguide with length of 13 $\rm cm$ for conventional SBS (without IDT) and  a length of 100 $ \rm\mu m$ in the present of IDT is equal to 2 $\rm \mu W$.  It can be seen that the Stokes is amplified exponentially through the backward direction while the pump is depleted in the forward direction.
 Although by applying the IDT the $\Upsilon_0$ coefficient will reduce drastically, however; the parameters of $\Upsilon_i$ (i=1,2,3) tabulated in table 2, play a key role in Stokes amplification as shown in \figref{ERSBS100}.
 As can be seen larger Stokes amplification in this situation occurs compared with conventional SBS process; this reduces the scale of the integrated Brillouin chip about two order of magnitude (from 13 $ \rm cm$ to 100 $\rm \mu m$ in this case).
 \begin{widetext}
	\begin{center}
		\begin{tabular}{c|c | c| c |c|c|c		}
		
			\multicolumn{7} { c  }{Table.2: The values of ERSBS coefficients of governing power equation for the proposed mode  }\\
   \\
				State & Angular frequency $\rm {[GRad/s]}$ & $\Upsilon_{\rm i,piezo}$ $\rm{[W^{-1}m^{-1}]}$
			& $\Upsilon_0$ $\rm{[W^{-1}m^{-1}]}$ & $\Upsilon_1$$\rm{[W^{-1}m^{-1}]}$ & $\Upsilon_2$$\rm{[W^{-2}m^{-1}]}$&
			$\Upsilon_3$$\rm{[m^{-1}]}$\\
			\hline
			ERSBS&66.05&-1.0293$\times$10$^{4}$& 	0.00302	&1.8947$\times$10$^{8}$ &	34.0976  &	-6.6448$\times$ 10$^{8}$\\
		
		\end{tabular}
	\end{center}
\end{widetext}

 \section{Conclusion}
We theoretically demonstrated Brillouin scattering in a piezoelectric medium in the presence of externally injection of acoustic phonons. We derived the governing equations corresponding to this  electromechanical interaction. It was shown how envelope of the excited acoustic mode is influenced by piezoelectricity and the intensity of the externally injected wave. We observed that the piezoelectric forces play a critical role in the optomechanical interactions; these  forces  appeared in the piezoelectric overlap integral ($\widetilde Q_{\rm i,piezo} (i=1,2)$) through the projection of the excited acoustic mode on the piezoelectric forces.
The power conversion between pump and Stokes in the piezoelectric medium was obtained and its characteristics were compared with the SBS governing equations in non-piezoelectric waveguides. As a case study, the method was investigated in GaAS nanowire. It was found that Stokes wave in the presence of applied piezoelectric power is strongly amplified in a straight piezoelectric waveguide of a few hundred microns length; to reach the same order of amplification in a silicon nanowire much longer(two orders of magnitude longer) length of waveguide is required.


\section { Appendix } 

\subsection { Wave interactions in the presence of an
externally injected piezoelectric signal}
By projecting $\widetilde{\phi}^*\times(-i\Omega)$ on \eq{11} ($\nabla \cdot\big(-\bm{\bm \epsilon}\cdot\nabla(\Psi\widetilde{\phi})+\mathbf e :\nabla (b\mathbf{\widetilde{u}})\big)=0$) and dropping the higher order derivatives of $\Psi$  we have
\begin{widetext}
	\begin{equation}
	\begin{split}
&	<(i\Omega\widetilde{\phi})| -\nabla \cdot(\bm{\bm \epsilon}\cdot\nabla(\Psi\widetilde{\phi}))+\nabla \cdot( \mathbf e :\nabla (b\mathbf{\widetilde{u}}))>\approx <(i\Omega\widetilde{\phi})| -\frac{\partial \Psi}{\partial z}\big(([\hat{a}_z]_{1\times3}\cdot (\bm{\bm \epsilon}\cdot\nabla\widetilde{\phi})+\nabla\cdot(\bm{\bm \epsilon}\cdot [\widetilde{\phi}]_{\rm z}))\big)\\&-\Psi\big(\nabla\cdot(\bm{\bm \epsilon}\cdot\nabla \widetilde{\phi})\big))+\frac{\partial b}{\partial z}\big(([\hat{a}_z]_{1\times3}\cdot (\mathbf e:\nabla\mathbf{\widetilde{u}}))+\nabla\cdot(\mathbf e: [\mathbf{\widetilde{u}}])\big)+b\big(\nabla\cdot(r:\nabla \mathbf{\widetilde{u}})\big)>=0,
	\label{projection_RF}
	\end{split}
	\end{equation}
\end{widetext}
\noindent where $[\hat{a}_z]_{1\times3}$ is the unit vector $[0 , 0, 1]$, $[\widetilde{\phi}]_{\rm z}$ is defined as $[0 , 0, \widetilde{\phi}]$  and $[\mathbf{\widetilde{u}}]$  denotes $[0 ,0,\mathbf{\widetilde{u}_z},\mathbf{\widetilde{u}_y},\mathbf{\widetilde{u}_x}, 0]^{\rm T}$.  Thus, the relation between the piezopotential envelope and its corresponding acoustic envelope takes the following form\\
\begin{equation}
\begin{split}
\frac{\partial  \Psi}{\partial z}{M_{\rm RF}^{(1)}}+\Psi {M_{\rm RF}^{(2)}}
=\frac{\partial b}{\partial z}{M_{\rm RF}^{(3)}}+ b {M_{\rm RF}^{(4)}},
\label{both_Psi-acoustic}
\end{split}
\end{equation}\\
\noindent where the coefficient ${M_{\rm RF}^{(i)}}$ (i =1,2,3,4) is an integral expressed by:
\begin{equation}
\begin{aligned}
& {M_{\rm RF}^{(1)}}=(-i\Omega)\int \widetilde{\phi}^{*}\big(([\hat{a}_z]_{1\times3}\cdot (\bm{\bm \epsilon}\cdot\nabla\widetilde{\phi})+\nabla\cdot(\bm{\bm \epsilon}\cdot [\widetilde{\phi}]_{\rm z})\big)ds,\\
& {M_{\rm RF}^{(2)}}=(-i\Omega)\int \widetilde{\phi}^{*}\big(\nabla\cdot(\bm{\bm \epsilon}\cdot\nabla \widetilde{\phi})\big))ds,\\
& {M_{\rm RF}^{(3)}}=(-i\Omega)\int \widetilde{\phi}^{*}\big(([\hat{a}_z]_{1\times3}\cdot (\mathbf  e
:\nabla\mathbf{\widetilde{u}}))+\nabla\cdot(\mathbf e: [\mathbf{\widetilde{u}}])\big) ds,\\
&{M_{\rm RF}^{(4)}}=(-i\Omega)\int \widetilde{\phi}^{*}\big(\nabla\cdot(\mathbf e:\nabla \mathbf{\widetilde{u}})\big)ds.
\end{aligned}
\end{equation}
then after a rearrangement, it is 
\begin{equation}
	\zeta_{\rm 0}\frac{\partial\Psi}{\partial z}+\alpha_{\rm ElM}\Psi=\frac{\partial b}{\partial z}+ \alpha_{\rm MEl}b,
\label{both_Psi-acoustic_appendix}
\end{equation}
where  the coefficients $	\zeta_{\rm 0}$, $\alpha_{\rm ElM}$ and $\alpha_{\rm MEl}$ are  $\frac{M_{\rm RF}^{(1)}}{M_{\rm RF}^{(3)}}$, $\frac{M_{\rm RF}^{(2)}}{M_{\rm RF}^{(3)}}$ and $\frac{M_{\rm RF}^{(4)}}{M_{\rm RF}^{(3)}}$, respectively.\\
{ By making an extra z-derivative  of the piezopotential envelope of \eq{both_Psi-acoustic_text}  and neglecting the second order derivatives of $b$ and $\Psi$, we have
\begin{equation}
\begin{split}
 \alpha_{\rm MEl}\frac{\partial b}{\partial z}\approx \alpha_{\rm ElM}\frac{\partial \Psi}{\partial z}.
\label{approx_final_dbPsi_text}
\end{split}
\end{equation}
Then, the piezopotential envelope ($\Psi$) can be obtained by substituting \eq{approx_final_dbPsi_text} in \eq{both_Psi-acoustic_text}
 \begin{equation}
\begin{split}
\Psi \approx \bigg( \frac{1-\zeta_{\rm 0}\frac{\alpha_{\rm MEl}}{\alpha_{\rm ElM}}}{\alpha_{\rm ElM}}\bigg)\frac{\partial b}{\partial z}+ \frac{\alpha_{\rm MEl}}{\alpha_{\rm ElM}}b.
\label{Psi-approx_b} 
\end{split}
\end{equation}}\\
\subsection{RF wave generated by optical interactions}
The interaction between pump and Stokes in a noncentro-symmetric material can generate RF wave due to diffrence frequency generation (DFG). Through the SBS process, only those RF signals that possess a frequency equal to the Brillouin frequency $\Omega$ can be engaged effectively in interactions. The electric displacement field in non-centrosymmetric material for this scenario can be expressed by \cite{lax1976electrodynamics,nelson1976linear}
\begin{equation}
\begin{split}
&\mathbf D_{\rm RF}={\bm{\bm \epsilon}}\cdot  \mathbf{E_{\rm RF}}+\mathbf e^{\rm F}: \mathbf S+\bm \epsilon_0\bm \chi^{(2)} \cdot \mathbf E_{\rm p}\mathbf E^{*}_{\rm s}+{\it c.c.},\\
\label{general_displacement_Strain}
\end{split}
\end{equation}
where $\mathbf e^{\rm F}$ is \cite{nelson1976linear}
\begin{equation}
\begin{split}
&\mathbf e^{\rm F}=\mathbf e-\bm \epsilon_0\bm \chi^{(2)} \cdot \mathbf E_{\rm p}\mathbf E^{*}_{\rm p}-\bm \epsilon_0\bm \chi^{(2)} \cdot \mathbf E_{\rm s}\mathbf E^{*}_{\rm s}\approx \mathbf e.
\label{general_displacement_Strain}
\end{split}
\end{equation}
Before investigating the contribution of second order nonlinearity $\bm \chi^{(2)}$, We review the phase matching condition in naoncentro-symmetric material when DFG take places. There are three wave numbers that play role in this situations; i.e. Pump and Stokes wavenumbers $k_{\rm p}$, $k_{\rm s}$ and and that of their nonlinear process, $k_{\rm DFG}$. Phase matching occurs when 
\begin{equation}
\Delta{k}=k_{\rm p}\pm k_{\rm s}-k_{\rm DFG}=0,
\label{dk}
\end{equation}
If we expand the wave numbers, we have
\begin{equation}
\Delta{k}=\frac{\omega_{\rm p} n(\omega_{\rm p})}{c}\pm \frac{\omega_{\rm s} n(\omega_{\rm s})}{c}-\frac{\Omega_{\rm RF} n(\Omega_{\rm RF})}{c}=0,
\label{dk}
\end{equation}
\noindent $\Delta k$ becomes nonzero in a non-centrosymmetric material through DFG process for backward scattering while under certain circumstances in forward scattering it can be nullified to achieve phase matching condition. A non zero  $\Delta k$ thus, can drastically reduce the power conversion between pump and Stokes, hence a weak RF envelope is generated through their interaction. 
Even with the assumption of phase matching condition, the value of $\bm \epsilon_0\bm \chi^{(2)} \cdot \mathbf E_{\rm p}\mathbf E^{*}_{\rm s}$ through the electrooptic (EO) effect is very small in comparison with the piezoelectric stress tensor term $\mathbf e: \mathbf S$ \cite{biaggio1999nonlocal}. 
With that, we can neglect the corresponding RF signal from DFG and in general EO effect.

\subsection { Acoustic wave in a piezoelectric Material} 
 
\noindent By a substitution \eq {U-centro} and \eq {T-centro} in the acoustic wave equation and neglecting higher order terms we have 

\begin{widetext}
	\begin{equation}
	\begin{split}
	&\frac{\partial b}{\partial z}\bigg[\bigg([\hat{a}_z]_{3\times6}\cdot (\mathbf c:\nabla\mathbf{\widetilde{u}})+\nabla\cdot(\mathbf c: [\mathbf{\widetilde{u}}])\bigg)
	+\bigg((-i\Omega)\big(([\hat{a}_z]_{3\times6}\cdot ( \bm\eta:\nabla\mathbf{\widetilde{u}}))+\nabla\cdot(\bm\eta: [\mathbf{\widetilde{u}}]))\big)\bigg)\bigg]\\&+b\bigg[\nabla\cdot(\mathbf c:\nabla \mathbf{\widetilde{u}})+\big((-i\Omega)\big(\nabla\cdot( \bm\eta:\nabla \mathbf{\widetilde{u}})\big)+\big(\rho\Omega^{2}\mathbf{ \widetilde u }\big)\bigg]+\frac{\partial \Psi}{\partial z}\mathbf {\widetilde f}_{\rm 1,piezo}+\Psi\mathbf {\widetilde f}_{\rm 2,piezo}+A_{\rm p}A^{*}_{\rm s} \mathbf {\widetilde f}_{\rm EM} +{\it c.c.}=0.
	\label{differential_final-acoustic2}
	\end{split}
	\end{equation}
\end{widetext}
where the matrix $[\hat{a}_z]_{3\times6}$ is
\begin{equation}
[\hat{a}_z]_{3\times6}=
\begin{bmatrix}
0 & 0 & 0&0& 1 & 0 \\
0 & 0 & 0&1 & 0&0 \\
0 & 0 & 1& 0 & 0&0 \\
\end{bmatrix}.
\end{equation} 
\noindent We now project the acoustic velocity mode $ \widetilde{\mathbf v}^{*} $ on \eq{differential_final-acoustic2} 
\newpage
\begin{widetext} 
	\begin{equation}
	\begin{split}
	&<\widetilde{\mathbf v}|\frac{\partial b}{\partial z}\bigg[\bigg([\hat{a}_z]_{3\times6}\cdot (\mathbf c:\nabla\mathbf{\widetilde{u}})+\nabla\cdot(\mathbf c: [\mathbf{\widetilde{u}}])\bigg)
	+\bigg((-i\Omega)\big(([\hat{a}_z]_{3\times6}\cdot ( \bm\eta:\nabla\mathbf{\widetilde{u}}))+\nabla\cdot( \bm\eta: [\mathbf{\widetilde{u}}]))\big)\bigg)\bigg]\\&+b\bigg[\nabla\cdot(\mathbf c:\nabla \mathbf{\widetilde{u}})+\big((-i\Omega)\big(\nabla\cdot( \bm\eta:\nabla \mathbf{\widetilde{u}})\big)+\big(\rho\Omega^{2}\mathbf{ \widetilde u }\big)\bigg]+\frac{\partial \Psi}{\partial z}\mathbf {\widetilde f}_{\rm 1,piezo}+\Psi\mathbf {\widetilde f}_{\rm 2,piezo}+A_{\rm p}A^{*}_{\rm s} \mathbf {\widetilde f}_{\rm EM}+{\it c.c.} >=0.
	\label{25}
	\end{split}
	\end{equation}
\end{widetext}

\noindent Then, by knowing that the eigenmode 
$\mathbf{\widetilde{u}} $ (i.e.  $b=1$) satisfies \cite{wolff2015stimulated}
\begin{equation}
\nabla\cdot(\mathbf c:\nabla \mathbf{\widetilde{u}})+\big(\rho\Omega^{2}\mathbf{ \widetilde u }\big)=0,
\end{equation}
\noindent the differential equation governing the variations of the acoustic envelope is given by
\begin{widetext}
	\begin{equation}
	\begin{split}
	\frac{\partial b}{\partial z}+\alpha_{\rm ac} b=\frac{-i\Omega \widetilde{Q}_{\rm b}}{\mathcal  P_{\rm ac}+i \mathcal P_{\rm LOSS}}A_{\rm p}A^{*}_{\rm s}+\frac{-i\Omega \widetilde Q_{\rm 1,piezo}}{\mathcal  P_{\rm ac}+i \mathcal P_{\rm LOSS}}\frac{\partial \Psi}{\partial z}+\frac{-i\Omega \widetilde Q_{\rm 2,piezo}}{\mathcal  P_{\rm ac}+i \mathcal P_{\rm LOSS}}\Psi,
	\label{final_acoustic_eq}
	\end{split}
	\end{equation}
	\end{widetext}
\noindent where $\alpha_{\rm ac}$, $\mathcal  P_{\rm ac}$, $\mathcal P_{\rm LOSS}$, $ \widetilde{Q}_{\rm i,piezo}$ (i=1,2) and $ \widetilde{Q}_{\rm b}$ are the linear loss coefficient, acoustic Poynting power, acoustic loss power and overlap integrals of piezoelectric and opto-acoustic, respectively. The acoustic loss coefficient $ \alpha_{\rm ac} $ is obtained by

	\begin{equation}	
	\begin{split}
		\alpha_{\rm ac}=\frac{\Omega^2}{(\mathcal P_{\rm ac}+i\mathcal P_{\rm LOSS})}\int \mathbf {\widetilde u}^{*} \cdot(\big(\nabla\cdot(\bm\eta:\nabla\mathbf{\widetilde{u}})\big) ds,
	\end{split}
	\end{equation}\\
\noindent and acoustic loss power is
	\begin{equation}	
	\begin{split}
	&\mathcal P_{\rm LOSS}={\Omega}\int  \mathbf {\widetilde u}^{*} \cdot
	\bigg((-i\Omega)\big(([\hat{a}_z]_{3\times6}\cdot ( \bm\eta:\nabla\mathbf{\widetilde{u}}))\\&+\nabla\cdot(\bm\eta: \mathbf{\widetilde{u}}))\big)\bigg)ds. 	
	\end{split}
	\end{equation}
The first  part of $\frac{\partial b}{\partial z}$ coefficient in \eq{25} can be denoted by the use of approximation applied in \cite{wolff2015stimulated} equalized to acoustic Poynting power ($\mathcal  P_{\rm ac}$) as shown in \eq{toatl_acoustic} (	$\mathcal  P^{\rm pure}_{\rm M}$). 
	\begin{equation}	
	\begin{split}
\mathcal  P_{\rm ac}\approx{i\Omega}\int  \mathbf {\widetilde u}^{*} \cdot\bigg([\hat{a}_z]_{3\times6}\cdot (\mathbf c:\nabla\mathbf{\widetilde{u}})+\nabla\cdot(\mathbf c: \mathbf{\widetilde{u}})\bigg) ds.
\label{250}\\
	\end{split}
	\end{equation}\\
\subsection{Energy density and power flow in a piezoelectric medium}
In general, the total energy of a propagating medium is summation of kinetic and stored (potential) energy\\
\begin{equation}
\begin{split}
&\mathcal W_{\rm tot}=\mathcal W_{\rm  kinetic}+\mathcal W_{\rm stored},
\label{energy_total}
\end{split}
\end{equation}\\
\noindent where kinetic energy is
\begin{equation}
\begin{split}
&\mathcal W_{\rm  kinetic}=-\frac{\Omega^{2}}{2}\int \rho ( \mathbf {\widetilde u}^{*}\cdot\mathbf {\widetilde u})dv +{\it c.c.},
\end{split}
\end{equation}\\
and  stored energy can be divided to mechanical and electrical energy.\\
\begin{equation}
\begin{split}
\mathcal W_{\rm stored}=\mathcal W^{\rm M}_{\rm tot}+\mathcal W^{\rm  El}_{\rm tot}.
\end{split}
\end{equation}
The total time-average density of  stored mechanical energy $\mathcal W^{\rm M}_{\rm tot}$ and electrical energy $\mathcal W^{\rm  El}_{\rm tot}$  can be expressed as \cite{auld1973acoustic} \\
\begin{equation}
\begin{split}
&\mathcal W^{\rm M}_{\rm tot}=\int(\mathbf T^{\rm piezo}_{\rm tot}  :\mathbf S^*)dv,\\
&\mathcal W^{\rm  El}_{\rm tot}=\int(\mathbf D_{\rm RF}^* \cdot\mathbf E_{\rm RF})dv,
\label{energy}
\end{split}
\end{equation}
\noindent where  $\mathbf T^{\rm piezo}_{\rm tot}=\mathbf c:\mathbf S-\mathbf e.\mathbf E_{\rm RF}$ is the  stress.
\noindent It must be noted that  the energy and power differs from Ref \cite{auld1973acoustic} by a factor of  $\frac{1}{2}$, is due to our convention used in \eq{U-centro}. 
\noindent By substitution of \eq{displacement_Strain} and \eq {T-centro} in \eq{energy}
we have
\begin{equation}
\begin{split}
&\mathcal W^{\rm M}_{\rm tot}=\int\{(\mathbf c:\mathbf S):\mathbf S^*-( \mathbf e.\mathbf E_{\rm RF}):\mathbf S^*\}dv\\&=\mathcal W^{\rm M}_{\rm pure}+\mathcal W^{\rm MEl},\\&
\mathcal W^{\rm  El}_{\rm tot}=\int\{({{\bm\epsilon}}\cdot  \mathbf{E_{\rm RF}})^* \cdot\mathbf E_{\rm RF}+(\mathbf e: \mathbf S)^* \cdot\mathbf E_{\rm RF}\}dv\\&=\mathcal W^{\rm  El}_{\rm pure}+\mathcal W^{\rm  ElM},\\
\label{energy_divided}
\end{split}
\end{equation}\\
where, \\
\begin{equation}
\begin{split}
&\mathcal W^{\rm  M}_{\rm pure}=\int\{( \mathbf c: \mathbf S):\mathbf S^*\}dv,\\&
\mathcal W^{\rm  El}_{\rm pure}=\int\{(\bm \epsilon.\mathbf E_{\rm RF})^*\cdot \mathbf E_{\rm RF}\}dv,\\&
\mathcal W^{\rm MEl}=\int\{-( \mathbf e.\mathbf E_{\rm RF}):\mathbf S^*\}dv,\\&
\mathcal W^{\rm  ElM}=\int\{(\mathbf e: \mathbf S)^*\cdot \mathbf E_{\rm RF}\}dv,\\
\end{split}
\end{equation}
and energy in terms of integral form is\\
\begin{equation}
\begin{split}
&\mathcal W^{\rm MEl}=\int( \mathbf e.\nabla(\Psi\widetilde{\phi}) ):\nabla(b\mathbf{\widetilde{u}})^{*}) dv+{\it c.c.},\\&\mathcal W^{\rm ElM}=-\int(( \mathbf e:\nabla(b\mathbf{\widetilde{u}}))^{*}\cdot\nabla(\Psi\widetilde{\phi}) ) dv+{\it c.c.}.
\end{split}
\end{equation}\\
It can be observed that mechanical energy   (electrical energy) are divided into pure mechanical (electrical) and mechanoelectrical (electromechanical) energy.
The densities of mechanoelectrical and electromechanical energies in arbitrary piezoelectric media are
always equal in absolute value and have opposite signs  \cite{zaitsev2003energy}. which means \\
\begin{equation}
\begin{split}
|\mathcal W^{\rm MEl}|=-|\mathcal W^{\rm  ElM}|.
\label{energy_divided2}
\end{split}
\end{equation}\\
Now the piezoelectric Poynting  vector  $\mathbf P_{\rm piezo}$ with quasi static approximation including acoustic and RF parts is calculated by \cite{auld1973acoustic}:\\
\begin{equation}
\begin{split}
 P_{\rm piezo} =\int(-{\mathbf V^*\cdot \mathbf T^{\rm piezo}_{\rm tot}}+{{\Phi}(-i\Omega\mathbf D_{\rm RF})^* })\cdot \mathbf ds+{\it c.c.},
\label{piezo_poyinting}
\end{split}
\end{equation}\\
\\
\\
and after an expansion, it is\\
\\
\\
\begin{widetext}
\begin{equation}
\begin{split}
& P_{\rm piezo}=|b|^2\bigg(\int\big(-\mathbf {\widetilde v}^*\cdot  (\mathbf c:\nabla\mathbf{\widetilde{u}})\big) \cdot \mathbf ds\bigg)+b^{*}\frac{\partial b}{\partial z}\bigg(\int(-\mathbf {\widetilde v}^*\cdot (\hat{a}_z\cdot (\mathbf c:\mathbf{\widetilde{u}}))) \cdot \mathbf ds\bigg)+b^{*}\Psi\bigg( \int(-\mathbf {\widetilde v}^*\cdot (\mathbf e\cdot\nabla\widetilde{\phi})) \cdot \mathbf ds\bigg)\\
&+b^{*}\frac{\partial \Psi}{\partial z}\bigg(\int(-\mathbf {\widetilde v}^*\cdot (\hat{a}_z\cdot (\bm e:\mathbf{\widetilde{\phi}}))) \cdot \mathbf ds\bigg)+|\Psi|^2\bigg(\int -i\Omega\widetilde{\phi}\big( \bm{\epsilon \cdot \nabla\widetilde{\phi} })^{*}\cdot \mathbf{ ds}\bigg)+\Psi\frac{\partial \Psi}{\partial z}^{*}\bigg(\int(\hat{a}_z\cdot ( \bm \epsilon:\widetilde{\phi})^{*}\cdot \mathbf{ds}\bigg)\\
&+\Psi b^{*}\bigg(\int(\mathbf{ r: \nabla\widetilde{u}})^*\cdot \mathbf{ds}\bigg)+\Psi \frac{\partial b}{\partial z}^{*}\bigg(\int(\mathbf{\hat{a}_z}\cdot (\mathbf{ r}:\mathbf{\widetilde{u}}))^{*}\cdot \mathbf {ds} \bigg)+{\it c.c.}.
\end{split}
\end{equation}
\end{widetext}
It is assumed that $b(\mathbf c:\nabla\mathbf{\widetilde{u}})\gg\frac{\partial b}{\partial z}(\hat{a}_z\cdot (\mathbf c:\mathbf{\widetilde{u}}))$. This approximation is also noted in \cite{wolff2015stimulated}, and subsequently $\Psi (\mathbf e\cdot\nabla\widetilde{\phi})\gg\frac{\partial \Psi}{\partial z}(\hat{a}_z\cdot (\mathbf e:\mathbf{\widetilde{\phi}})) $. Therefore, we neglect it in calculating the piezoelectric power.
\begin{widetext}
\begin{equation}
\begin{split}
 & P_{\rm piezo}\approx|b|^2\bigg(\int\big(-\mathbf {\widetilde v}^*\cdot  (\mathbf c:\nabla\mathbf{\widetilde{u}})\big) \cdot \mathbf ds\bigg)+b^{*}\Psi\bigg( \int(-\mathbf {\widetilde v}^*\cdot (\mathbf e\cdot\nabla\widetilde{\phi})) \cdot \mathbf ds\bigg)
+|\Psi|^2\bigg(\int\bigg(-i\Omega{\widetilde{\phi}}\big( ((\bm \epsilon \cdot \nabla\widetilde{\phi} )^{*}\cdot \mathbf ds\bigg)\\&+\Psi b^{*}\bigg(\int(-i\Omega{\widetilde{\phi}\big(\bf r: \nabla\widetilde{u})^* })\cdot \mathbf ds\bigg)+{\it c.c.}
\end{split}
\end{equation}
\end{widetext}
\noindent By substituting \eq{approx_final_dbPsi_text} and \eq{Psi-approx_b} we have
\begin{widetext}
\begin{equation}
\begin{split}	
&P_{\rm piezo}\approx|b|^2\bigg(\int\big(-\mathbf {\widetilde v}^*\cdot  (\mathbf c:\nabla\mathbf{\widetilde{u}})\big) \cdot \mathbf ds\bigg)+b^{*}\bigg\{{\frac{\partial b}{\partial z}\big( \frac{1-\zeta_{\rm 0}\frac{\alpha_{\rm MEl}}{\alpha_{\rm ElM}}}{\alpha_{\rm ElM}}\big)+ b\frac{\alpha_{\rm MEl}}{\alpha_{\rm ElM}}\bigg\}}\\
&\bigg( \int(-\mathbf {\widetilde v}^*\cdot (\mathbf e\cdot\nabla\widetilde{\phi})) \cdot \mathbf ds\bigg)	
+\bigg|\bigg\{{\frac{\partial b}{\partial z}\big( \frac{1-\zeta_{\rm 0}\frac{\alpha_{\rm MEl}}{\alpha_{\rm ElM}}}{\alpha_{\rm ElM}}\big)+ b\frac{\alpha_{\rm MEl}}{\alpha_{\rm ElM}}\bigg\}}\bigg|^2\\
&\bigg(\int\bigg(-i\Omega{\widetilde{\phi}\big( ((\bm \epsilon \cdot \nabla\widetilde{\phi} )^{*}\cdot \mathbf ds\bigg)+ \bigg\{{\frac{\partial b}{\partial z}\big( \frac{1-\zeta_{\rm 0}\frac{\alpha_{\rm MEl}}{\alpha_{\rm ElM}}}{\alpha_{\rm ElM}}\big)+ b\frac{\alpha_{\rm MEl}}{\alpha_{\rm ElM}}\bigg\}} b^{*}\bigg(\int(-i\Omega{\widetilde{\phi}}\big(\bf r: \nabla\widetilde{u})^* }\cdot \mathbf ds\bigg)+{\it c.c.}.
\end{split}
\end{equation}
\end{widetext}
Here, by neglecting the differential terms, the power can be simplified as follows. This approximation is valid as long as $b$ has a slowly varying envelop and $\Psi$ is equal to $\frac{\alpha_{\rm MEl}}{\alpha_{\rm ElM}}b$ according to  \eq{Psi-approx_b}.
\begin{widetext}
\begin{equation}
\begin{split}
 P_{\rm piezo}\approx|b|^2 (\mathcal P^{\rm M}_{\rm pure}+\mathcal P^{\rm MEL}+ \mathcal P^{\rm El}_{\rm pure}+\mathcal P^{\rm ELM})=|b|^2 (\mathcal  P^{\rm M}_{\rm tot}+\mathcal  P^{\rm El}_{\rm tot})=|b|^2 \mathcal P_{\rm piezo},
 \label{piezo-power}
\end{split}
\end{equation}
\end{widetext}
\noindent where $\mathcal  P^{\rm M}_{\rm tot}$ is acoustic power in a non-centrosymmetric material  and can be expressed by
\begin{equation}
\begin{split}
&\mathcal  P^{\rm M}_{\rm tot}={-i\Omega}\int  \mathbf {\widetilde u}^{*} \cdot\bigg( (\mathbf c:\nabla\mathbf{\widetilde{u}}+\frac{\alpha_{\rm MEl}}{\alpha_{\rm ElM}} (\mathbf e\cdot\nabla\widetilde{\phi})\bigg)\cdot \mathbf ds+{\it c.c.}\\&=
\mathcal  P^{\rm M}_{\rm pure}+\mathcal  P^{\rm MEl},
\label{toatl_acoustic}
\end{split}
\end{equation}
\noindent where $\mathcal  P^{\rm M}_{\rm pure}$ and $\mathcal  P^{\rm MEl}$ are pure mechanical power and mechanoelectrical power, respectively. It must be noted that pure mechanical power $\mathcal  P^{\rm M}_{\rm pure}$ is the same with $\mathcal  P_{\rm ac}$ ($\mathcal  P^{\rm M}_{\rm pure}$ = $\mathcal  P_{\rm ac}$). And they can be written as follows:
\begin{equation}
\begin{split}
&\mathcal  P^{\rm M}_{\rm pure}={-i\Omega}\int  \mathbf {\widetilde u}^{*} \cdot (\mathbf c:\nabla\mathbf{\widetilde{u}})\cdot \mathbf ds+{\it c.c.},\\&
\mathcal  P^{\rm MEl}=
{-i\Omega}\frac{\alpha_{\rm MEl}}{\alpha_{\rm ElM}}\int  \mathbf {\widetilde u}^{*} \cdot (\mathbf e\cdot\nabla\widetilde{\phi})\cdot \mathbf ds+{\it c.c.},
\end{split}
\end{equation}
\noindent and $\mathcal  P^{\rm El}_{\rm tot}$ is the total electrical part of the Poynting  power in a piezoelectric medium, which can itself be divided to pure electrical and electromechanical power.
\begin{equation}
\begin{split}
&\mathcal  P^{\rm El}_{\rm tot}={-i\Omega}|\frac{\alpha_{\rm MEl}}{\alpha_{\rm ElM}}|^2\int\widetilde{\phi}(\bm \epsilon \cdot \nabla\widetilde{\phi} )^{*}\cdot\mathbf {ds} \\&-\frac{\alpha_{\rm MEl}}{\alpha_{\rm ElM}} {i\Omega}\int\widetilde{\phi}(\bf r: \nabla\widetilde{u} )^{*}\cdot \mathbf ds+{\it c.c.}=\mathcal  P^{\rm El}_{\rm pure}+\mathcal P^{\rm {ElM}},
\label{P_{RF}}
\end{split}
\end{equation}
\noindent where 
\begin{equation}
\begin{split}
&\mathcal  P^{\rm El}_{\rm pure}={-i\Omega}|\frac{\alpha_{\rm MEl}}{\alpha_{\rm ElM}}|^2\int\widetilde{\phi}(\bf \bm \epsilon \cdot \nabla\widetilde{\phi} )^{*}\cdot\mathbf {ds}+{\it c.c.},\\&
\mathcal P^{\rm {ElM}}={-i\Omega}\frac{\alpha_{\rm MEl}}{\alpha_{\rm ElM}}\int\widetilde{\phi}(\bf r: \nabla\widetilde{u} )^{*}\cdot \mathbf ds +{\it c.c.}.
\end{split}
\end{equation}
\noindent It should be noted that for many piezoelectric materials such as GaAs or AlN, the relation ${M_{\rm RF}^{(1)}} =|\frac{{M_{\rm RF}^{(2)}}}{M_{\rm RF}^{(4)}}|^2\mathcal P^{\rm El}_{\rm pure}$ occurs due to  their isotropic material properties (i.e. having diagonal permittivity tensor ($\bm \epsilon$)).
Also, for Bulky cubic crystal structure the electromechanical power is equal to absolute value of the mechanoelectrical power because the tensor of $\mathbf e_{ijk}=\mathbf e_{kij}$ \cite{zaitsev2003energy}, however; in an integrated the mode profile can play an important role. Therefore, in nano waveguide structure even with cubic crystal material the electromechanical power is not the same absolute value with the mechanoelectrical power.
\subsection { Deriving the Power Conversion Equations}
We begin with \eq{both_Psi-acoustic_appendix} and after a rearrangement it is  
	\begin{equation}
	\begin{split}
\frac{\partial b}{\partial z}=- \alpha_{\rm MEl}b +\zeta_0\frac{\partial  \Psi}{\partial z}+\alpha_{\rm ElM}\Psi .
\end{split}
\end{equation}\\
\noindent Then, by substituting in \eq {final_acoustic_eq}, the  acoustic envelope is
\begin{equation}
	\begin{split}
&b=\gamma_1 A_{\rm p}A^{*}_{\rm s}+\gamma_2\frac{\partial \Psi}{\partial z}+\gamma_3\Psi,
\label{exact_bb}
\end{split}
	\end{equation}
where $\gamma_i$ (i=1,2,3) are coefficient as shown in \eq {gamma_i_text1}.\\ By substitution \eq{exact_bb} in \eq{pAp} and \eq{pAs} we have\\
\\
\begin{widetext}
	\begin{equation}
	\begin{split}
&\frac{\partial{A_{\rm p}}}{\partial{z}}=-\frac{i\omega_{\rm p} \widetilde Q_{\rm p}}{\mathcal P_{\rm p}}bA_{\rm s}=-\frac{i\omega_{\rm p} \widetilde Q_{\rm p}}{\mathcal P_{\rm p}}\bigg[\gamma_1 A_{\rm p}A^{*}_{\rm s}+\gamma_2\frac{\partial \Psi}{\partial z}+\gamma_3\Psi\bigg]A_{\rm s},
\label{pAp}
\end{split}
\end{equation}
\begin{equation}
	\begin{split}
&\frac{\partial{A_{\rm s}}}{\partial{z}}=-\frac{i\omega_{\rm s} \widetilde Q_{\rm s}}{\mathcal P_{\rm s}}A_{\rm p}b^*=-\frac{i\omega_{\rm s} \widetilde Q_{\rm s}}{\mathcal P_{\rm  s}}\bigg[\gamma_1 A_{\rm p}A^{*}_{\rm s}+\gamma_2\frac{\partial \Psi}{\partial z}+\gamma_3\Psi\bigg]^*A_{\rm p},
\label{pAs}
\end{split}
\end{equation}
\end{widetext}
where $P_{\rm i}=(-1)^{\xi}\mathcal P_{\rm i}|A_{\rm i}|^{2}$ in which $\xi$ is equal to  zero and one for forward and backward waves, respectively. From \eq{piezo-power} the power differential equation in a lossless medium can be written as 
\begin{widetext}
	\begin{align}
&\frac{\partial{P_{\rm p}}}{\partial{z}}=2\Re\{{\frac{-\Omega\omega_{\rm p} |\widetilde Q_{\rm b}|^{2}}{(\alpha_{\rm ac}-\alpha_{\rm MEl})\mathcal  P_{\rm ac}\mathcal P_{\rm p}}\}}\mathcal P_{\rm p} |A_{\rm p}|^{2}|A_{\rm s}|^{2}+2\Re\{{-\frac{i\omega_{\rm p} \widetilde Q_{\rm p}}{\mathcal P_{\rm p}}\gamma_2\mathcal P_{\rm p} A_{\rm p}^{*}A_{\rm s}\frac{\partial \Psi}{\partial z}\}}+2\Re\{{-\frac{i\omega_{\rm p} \widetilde Q_{\rm p}}{\mathcal P_{\rm p}}\gamma_3\mathcal P_{\rm p} A_{\rm p}^{*}A_{\rm s}\Psi\}},\\
&\frac{\partial{P_{\rm s}}}{\partial{z}}=\frac{2\Omega\omega_{\rm s} |\widetilde Q_{\rm b}|^{2}}{(\alpha_{\rm ac}-\alpha_{\rm MEl})\mathcal  P_{\rm ac}\mathcal P_{\rm s}}\mathcal P_{\rm s} |A_{\rm p}|^{2}|A_{\rm s}|^{2}+2\Re\{{(-\frac{i\omega_{\rm s} \widetilde Q_{\rm s}}{\mathcal P_{\rm s}})\bigg(\gamma_2\mathcal P_{\rm s} A_{\rm p}^{*}A_{\rm s}\frac{\partial \Psi}{\partial z}\bigg)^{*}\}}+2\Re\{{(-\frac{i\omega_{\rm s} \widetilde Q_{\rm s}}{\mathcal P_{\rm s}})\bigg(\gamma_3\mathcal P_{\rm s} A_{\rm p}^{*}A_{\rm s} \Psi\bigg)^{*}\}},\\
&\frac{\partial{P_{\rm piezo}}}{\partial{z}}=\mathcal P_{\rm piezo}(b\frac{\partial b^{*}}{\partial z}+b^{*}\frac{\partial b}{\partial{z}})=2{\mathcal P_{\rm piezo}\Re\{b^{*}\frac{\partial b}{\partial z}\}}.
\label{power}
	\end{align}
\end{widetext}
As shown in the appendix F, the two terms of  $A_{\rm p}^{*}A_{\rm s}\frac{\partial \Psi}{\partial z}$ and $A_{\rm p}^{*}A_{\rm s}\Psi$ can be written as
\begin{widetext}
 \begin{equation}
\begin{split}
&A^{*}_{\rm p}A_{\rm s}\Psi \approx\sigma_1  P_{\rm piezo}{P_{\rm p}}-\sigma_1  P_{\rm piezo}{P_{\rm s}}+\sigma_2  P_{\rm piezo}{P^{2}_{\rm p}}+\sigma_2  P_{\rm piezo}{P^{2}_{\rm s}}-2\sigma_2  P_{\rm piezo}{P_{\rm p}}{P_{\rm s}}+\sigma_3  P_{\rm piezo},\\
&A^{*}_{\rm p}A_{\rm s}\frac{\partial{\Psi}}{\partial{z}} \approx\tau_1  P_{\rm piezo}{P_{\rm p}}-\tau_1  P_{\rm piezo}{P_{\rm s}}+\tau_2  P_{\rm piezo}{P^{2}_{\rm p}}+\tau_2  P_{\rm piezo}{P^{2}_{\rm s}}-2\tau_2  P_{\rm piezo}{P_{\rm p}}{P_{\rm s}},
\end{split}
\end{equation}
\end{widetext}
where $\sigma_i$ and $\tau_i$ are piezo-coefficients.
\noindent Therefore, the power conversion differential equation can be expressed as follows 
 \begin{widetext}
	\begin{equation}
	\begin{split}
&\frac{\partial{P_{\rm p}}}{\partial{z}}=-\Upsilon_{\rm 0} P_{\rm p}P_{\rm s}-\Upsilon_{\rm 1}  P_{\rm piezo}{P_{\rm p}}+\Upsilon_{\rm 1} P_{\rm piezo}{P_{\rm s}}-\Upsilon_{\rm 2}  P_{\rm piezo}{P^{2}_{\rm p}}-\Upsilon_{\rm 2}  P_{\rm piezo}{P^{2}_{\rm s}}+2\Upsilon_{\rm 2}  P_{\rm piezo}{P_{\rm p}}{P_{\rm s}}-\Upsilon_{\rm 3}  P_{\rm piezo},\\
&\frac{\partial{P_{\rm s}}}{\partial{z}}=\Upsilon_{\rm 0} P_{\rm p}P_{\rm s}+\Upsilon_{\rm 1}  P_{\rm piezo}{P_{\rm p}}-\Upsilon_{\rm 1}  P_{\rm piezo}{P_{\rm s}}+\Upsilon_{\rm 2}   P_{\rm piezo}{P^{2}_{\rm p}}+\Upsilon_{\rm 2}  P_{\rm piezo}{P^{2}_{\rm s}}-2\Upsilon_{\rm 2}  P_{\rm piezo}{P_{\rm p}}{P_{\rm s}}+\Upsilon_{\rm 3}  P_{\rm piezo},\\
&\frac{\partial{P_{\rm piezo}}}{\partial{z}}=\Upsilon_{\rm piezo} P_{\rm piezo}( P_{\rm p}- P_{\rm s}),
\label{p_b}
	\end{split}
	\end{equation}
\end{widetext}
\noindent where ERSBS (EBS) coefficients $\Upsilon_0$ and  $\Upsilon_i$  (i=1:3) are 
	\begin{equation}
	\begin{split}
 &\Upsilon_0=2\Re\{{\frac{\Omega\omega_{\rm p} |\widetilde Q_{\rm b}|^{2}}{(\alpha_{\rm ac}-\alpha_{\rm MEl})\mathcal  P_{\rm ac}\mathcal P_{\rm p}\mathcal P_{\rm s}}\}}\label{modified_SBS_gain},
 	\end{split}
	\end{equation}
 \begin{equation}
	\begin{split}
 &\Upsilon_{\rm i}=\Upsilon_{\rm i,p}=2\Re\{{(\frac{-i\omega_{\rm p} \widetilde{Q}_{\rm p}}{\mathcal P_{\rm p}})\tau_i\gamma_2\}}+2\Re\{{(\frac{-i\omega_{\rm p} \widetilde{Q}_{\rm p}}{\mathcal P_{\rm p}})\sigma_i\gamma_3\}},
	\end{split}
	\end{equation}

\begin{equation}
\begin{split}
 &\Upsilon_{\rm i,s}=2\Re\{{(-\frac{i\omega_{\rm s} \widetilde{Q}_{\rm s}}{\mathcal P_{\rm s}})\tau_i^{*}\gamma_2^{*}\}}+2\Re\{{(\frac{-i\omega_{\rm s} \widetilde{Q}_{\rm s}}{\mathcal P_{\rm s}})\sigma_i^{*}\gamma_3^{*}\}}\\ &=2\Re\{{(\frac{i\omega_{\rm p} \widetilde{Q}_{\rm p}}{\mathcal P_{\rm s}})^{*}\tau_i^{*}\gamma_2^{*}\}}+2\Re\{{(\frac{i\omega_{\rm p} \widetilde{Q}_{\rm p}}{\mathcal P_{\rm s}})^{*}\sigma_i^{*}\gamma_3^{*}\}}=-\Upsilon_{\rm i},
	\end{split}
	\end{equation}
and piezo power loss ($\Upsilon_{\rm piezo}$) is
	\begin{equation}
		\begin{split}
\Upsilon_{\rm piezo}=2\Re\{{\gamma_4\}}.
\end{split}
\end{equation}
\subsection { Calculation of $A_{\rm p}^{*}A_{\rm s}\frac{\partial \Psi}{\partial z}$ and $A_{\rm p}^{*}A_{\rm s}\Psi$ Terms} 

\noindent By making an extra spacial derivative -along the propagation direction - of the acoustic envelope \eq {exact_bb}  and neglecting the second order derivative of the envelope (i.e.$\frac{\partial^2 \Psi}{\partial z^2}\approx\frac{\partial^2 b}{\partial z^2}\approx0$), we reach the following relation
	\begin{equation}
	\begin{split}
&\frac{\partial b}{\partial z}=\gamma_1\frac{\partial (A_{\rm p}A^{*}_{\rm s})}{\partial z}+\gamma_2\frac{\partial^{2} \Psi}{\partial z^{2}}+\gamma_3\frac{\partial \Psi}{\partial z}\\&\approx\gamma_1\frac{\partial (A_{\rm p}A^{*}_{\rm s})}{\partial z}+\gamma_3\frac{\partial \Psi}{\partial z},
\label{exact_b}
\end{split}
	\end{equation}
\noindent By substituting \eq{approx_final_dbPsi_text} in \eq{exact_b} and after a  rearrangement and then  by substituting  \eq{pAp} and \eq{pAs} for the pump and Stokes derivatives and by knowing that $P_{\rm i}=(-1)^{\xi}\mathcal P_{\rm i}|A_{\rm i}|^{2}$, the differential acoustic envelope variation in a relation with pump and Stokes envelope is expressed by\\
\\
\begin{widetext}
	\begin{equation}
	\begin{split}
&\frac{\partial b}{\partial z}\approx\big(\frac{\gamma_1}{1-\gamma_3\frac{\alpha_{\rm MEl}}{\alpha_{\rm ElM}}} \big)\frac{\partial (A_{\rm p}A^{*}_{\rm s})}{\partial z}=\big(\frac{\gamma_1}{1-\gamma_3\frac{\alpha_{\rm MEl}}{\alpha_{\rm ElM}}} \big) A^{*}_{\rm s}\frac{\partial A_{\rm p}}{\partial z}+\frac{\gamma_1}{1-\gamma_3\frac{\alpha_{\rm MEl}}{\alpha_{\rm ElM}}} A_{\rm p} \frac{\partial A^{*}_{\rm s}}{\partial z}\\&=\big(\frac{\gamma_1}{1-\gamma_3\frac{\alpha_{\rm MEl}}{\alpha_{\rm ElM}}} \big)(\frac{-i\omega_{\rm p} \widetilde{Q}_{\rm p}}{\mathcal P_{\rm p}})|A_{\rm s}|^{2}b+\big(\frac{\gamma_1}{1-\gamma_3\frac{\alpha_{\rm MEl}}{\alpha_{\rm ElM}}} \big) (\frac{i\omega_{\rm s} \widetilde{Q}^{*}_{\rm s}}{\mathcal P_{\rm s}})|A_{\rm p}|^{2}b\approx\gamma_4( P_{\rm p}- P_{\rm s})b,\\
\label{exact_db_gamma_4}
\end{split}
	\end{equation}
	\end{widetext}
 \noindent where $\gamma_4 $ is 
		\begin{equation}
	\begin{split}
\gamma_4=\big(\frac{\gamma_1}{1-\gamma_3\frac{\alpha_{\rm MEl}}{\alpha_{\rm ElM}}} \big)(\frac{i\omega_{\rm p} \widetilde{Q}_{\rm p}}{\mathcal P_{\rm p}\mathcal P_{\rm s}}).
\end{split}
	\end{equation}	
\noindent Therefore, $\frac{\partial \Psi}{\partial z}$ can be calculated by $\frac{\alpha_{\rm MEl}}{\alpha_{\rm ElM}}\frac{\partial b}{\partial z}$  ($\frac{\partial \Psi}{\partial z}\approx\frac{\alpha_{\rm MEl}}{\alpha_{\rm ElM}}\frac{\partial b}{\partial z}$). And after replacing \eq{exact_db_gamma_4} in \eq{Psi-approx_b}, equation of $\Psi$  can be expressed as follows\\
 \begin{equation}
\begin{split}
&\Psi \approx
\bigg(\gamma_4 \big( \frac{1-\zeta_{\rm 0}\frac{\alpha_{\rm MEl}}{\alpha_{\rm ElM}}}{\alpha_{\rm ElM}}\big)\bigg)( P_{\rm p}- P_{\rm s})b+\frac{\alpha_{\rm MEl}}{\alpha_{\rm ElM}}b.\\
\label{Psi-approx}
\end{split}
\end{equation}
Therefore, terms of $\Psi$ and $\frac{\partial \Psi}{\partial z}$ 
are calculated based on pump, Stokes and acoustic envelopes. In continue, the terms of $A_{\rm p}A^{*}_{\rm s}$ is realized. It is started with \eq{exact_bb} and then by a substitution of  \eq{exact_db_gamma_4} and \eq {Psi-approx} , we have\\
\begin{widetext}
	\begin{equation}
	\begin{split}
	&b=\gamma_1A_{\rm p}A^{*}_{\rm s}+\gamma_2\frac{\partial \Psi}{\partial z}+\gamma_3\Psi\\&=\gamma_1A_{\rm p}A^{*}_{\rm s}+\gamma_2\gamma_4\frac{\alpha_{\rm MEl}}{\alpha_{\rm ElM}}( P_{\rm p}- P_{\rm s}) b+\gamma_3\gamma_4 \big( \frac{1-\zeta_{\rm 0}\frac{\alpha_{\rm MEl}}{\alpha_{\rm ElM}}}{\alpha_{\rm ElM}}\big)( P_{\rm p}- P_{\rm s})b+\gamma_3\frac{\alpha_{\rm MEl}}{\alpha_{\rm ElM}}b,\\
\label{ApAS}
\end{split}
	\end{equation}
	\end{widetext}
\noindent by a rearrangement, it is
\begin{widetext}
	\begin{equation}
	\begin{split}
		&A_{\rm p}A^{*}_{\rm s}=(\frac{1}{\gamma_1}-\frac{\gamma_3}{\gamma_1}\frac{\alpha_{\rm MEl}}{\alpha_{\rm ElM}})b-\bigg(\frac{\gamma_2\gamma_4}{\gamma_1}\frac{\alpha_{\rm MEl}}{\alpha_{\rm ElM}}+\frac{\gamma_3\gamma_4}{\gamma_1} \big( \frac{1-\zeta_{\rm 0}\frac{\alpha_{\rm MEl}}{\alpha_{\rm ElM}}}{\alpha_{\rm ElM}}\big)\bigg) (P_{\rm p}- P_{\rm s})b,
\end{split}
	\end{equation}
	\end{widetext}
and then, we have
\begin{widetext}
 \begin{equation}
\begin{split}
&A^{*}_{\rm p}A_{\rm s}\Psi \approx\bigg((\frac{1}{\gamma_1}-\frac{\gamma_3}{\gamma_1}\frac{\alpha_{\rm MEl}}{\alpha_{\rm ElM}})b-\bigg(\frac{\gamma_2\gamma_4}{\gamma_1}\frac{\alpha_{\rm MEl}}{\alpha_{\rm ElM}}+\frac{\gamma_3\gamma_4}{\gamma_1} \big( \frac{1-\zeta_{\rm 0}\frac{\alpha_{\rm MEl}}{\alpha_{\rm ElM}}}{\alpha_{\rm ElM}}\big)\bigg) (P_{\rm p}- P_{\rm s})b\bigg)^{*}\\&\bigg(\big(\gamma_4 \big( \frac{1-\zeta_{\rm 0}\frac{\alpha_{\rm MEl}}{\alpha_{\rm ElM}}}{\alpha_{\rm ElM}}\big)\big)( P_{\rm p}- P_{\rm s})b+\frac{\alpha_{\rm MEl}}{\alpha_{\rm ElM}}b\bigg)\\&=\sigma_1  P_{\rm piezo}{P_{\rm p}}-\sigma_1  P_{\rm piezo}{P_{\rm s}}+\sigma_2  P_{\rm piezo}{P^{2}_{\rm p}}+\sigma_2  P_{\rm piezo}{P^{2}_{\rm s}}-2\sigma_2  P_{\rm piezo}{P_{\rm p}}{P_{\rm s}}+\sigma_3  P_{\rm piezo},\\
\end{split}
\end{equation}
\end{widetext}
\noindent where $\sigma_i$ are
\begin{widetext}
 \begin{equation}
\begin{split}
&\sigma_1=\frac{1}{\mathcal  P_{\rm piezo}}\bigg(-\frac{\alpha_{\rm MEl}}{\alpha_{\rm ElM}}\big(\frac{\gamma_2\gamma_4}{\gamma_1}\frac{\alpha_{\rm MEl}}{\alpha_{\rm ElM}}+\frac{\gamma_3\gamma_4}{\gamma_1} \big( \frac{1-\zeta_{\rm 0}\frac{\alpha_{\rm MEl}}{\alpha_{\rm ElM}}}{\alpha_{\rm ElM}}\big)\big)^{*}+\big(\gamma_4 \big( \frac{1-\zeta_{\rm 0}\frac{\alpha_{\rm MEl}}{\alpha_{\rm ElM}}}{\alpha_{\rm ElM}}\big)(\frac{1}{\gamma_1}-\frac{\gamma_3}{\gamma_1}\frac{\alpha_{\rm MEl}}{\alpha_{\rm ElM}})^{*}\big)\bigg)
,\\&\sigma_2=\frac{-1}{\mathcal  P_{\rm piezo}}\bigg(\frac{\gamma_2\gamma_4}{\gamma_1}\frac{\alpha_{\rm MEl}}{\alpha_{\rm ElM}}+\frac{\gamma_3\gamma_4}{\gamma_1} \big( \frac{1-\zeta_{\rm 0}\frac{\alpha_{\rm MEl}}{\alpha_{\rm ElM}}}{\alpha_{\rm ElM}}\big)\bigg)^{*}\bigg(\gamma_4 \big( \frac{1-\zeta_{\rm 0}\frac{\alpha_{\rm MEl}}{\alpha_{\rm ElM}}}{\alpha_{\rm ElM}}\big)\bigg)
,\\&\sigma_3=\frac{1}{\mathcal  P_{\rm piezo}}\frac{\alpha_{\rm MEl}}{\alpha_{\rm ElM}}(\frac{1}{\gamma_1}-\frac{\gamma_3}{\gamma_1}\frac{\alpha_{\rm MEl}}{\alpha_{\rm ElM}})^{*},
\end{split}
\end{equation}
\end{widetext}
and the other term is 
\begin{widetext}
\begin{equation}
\begin{split}
&A^{*}_{\rm p}A_{\rm s}\frac{\partial \Psi}{\partial z} \approx\bigg((\frac{1}{\gamma_1}-\frac{\gamma_3}{\gamma_1}\frac{\alpha_{\rm MEl}}{\alpha_{\rm ElM}})b-\bigg(\frac{\gamma_2\gamma_4}{\gamma_1}\frac{\alpha_{\rm MEl}}{\alpha_{\rm ElM}}+\frac{\gamma_3\gamma_4}{\gamma_1} \big( \frac{1-\zeta_{\rm 0}\frac{\alpha_{\rm MEl}}{\alpha_{\rm ElM}}}{\alpha_{\rm ElM}}\big)\bigg) (P_{\rm p}- P_{\rm s})b\bigg)^{*}(\gamma_4\frac{\alpha_{\rm MEl}}{\alpha_{\rm ElM}}( P_{\rm p}- P_{\rm s})b)\\&
=\tau_1  P_{\rm piezo}{P_{\rm p}}-\tau_1  P_{\rm piezo}{P_{\rm s}}+\tau_2  P_{\rm piezo}{P^{2}_{\rm p}}+\tau_2 P_{\rm piezo}{P^{2}_{\rm s}}-2\tau_2  P_{\rm piezo}{P_{\rm p}}{P_{\rm s}},
\end{split}
\end{equation}
\end{widetext}
 where $\tau_i$ are
\begin{widetext}
\begin{equation}
\begin{split}
&\tau_1=\frac{1}{\mathcal  P_{\rm piezo}}\big(\frac{1}{\gamma_1}-\frac{\gamma_3}{\gamma_1}\frac{\alpha_{\rm MEl}}{\alpha_{\rm ElM}}\big)^{*}(\gamma_4\frac{\alpha_{\rm MEl}}{\alpha_{\rm ElM}}),\\
&\tau_2=\frac{1}{\mathcal  P_{\rm piezo}}\bigg(-\bigg(\frac{\gamma_2\gamma_4}{\gamma_1}\frac{\alpha_{\rm MEl}}{\alpha_{\rm ElM}}+\frac{\gamma_3\gamma_4}{\gamma_1} \big( \frac{1-\zeta_{\rm 0}\frac{\alpha_{\rm MEl}}{\alpha_{\rm ElM}}}{\alpha_{\rm ElM}}\big)\bigg) \bigg)^{*}(\gamma_4\frac{\alpha_{\rm MEl}}{\alpha_{\rm ElM}}),
\end{split}
\end{equation}
\end{widetext}
\subsection {Closed form of Acoustic wave}
 The differential equation of the acoustic envelope is obtained by substituting \eq{approx_final_dbPsi_text} and (\ref{Psi-approx_b})  in \eq{final_acoustic_eq22}
\\
\begin{widetext}
	\begin{equation}
	\begin{split}
	\frac{\partial b}{\partial z}+\alpha_{\rm new}b=\frac{-i\Omega Q_{\rm b}}{\mathcal P_{\rm ac}+i\Omega \widetilde Q_{\rm 1,piezo}\frac{\alpha_{\rm MEl}}{\alpha_{\rm ElM}}+i\Omega \widetilde Q_{\rm 2,piezo}\big( \frac{1-\zeta_{\rm 0}\frac{\alpha_{\rm MEl}}{\alpha_{\rm ElM}}}{\alpha_{\rm ElM}}\big)}  A_{\rm p}A^{*}_{\rm s},\\
 \alpha_{\rm new}=\frac{\mathcal  P_{\rm ac}\alpha_{\rm ac}+\frac{\alpha_{\rm MEl}}{\alpha_{\rm ElM}}i\Omega \widetilde Q_{\rm 2,piezo}}{\mathcal P_{\rm ac}+i\Omega \widetilde Q_{\rm 1,piezo}\frac{\alpha_{\rm MEl}}{\alpha_{\rm ElM}}+i\Omega \widetilde Q_{\rm 2,piezo}\big( \frac{1-\zeta_{\rm 0}\frac{\alpha_{\rm MEl}}{\alpha_{\rm ElM}}}{\alpha_{\rm ElM}}\big)}. \\
 \label{compact}
		\end{split}
	\end{equation}
\end{widetext}
 The new formulation in \eq{compact} is interesting as it represents a the impact of final acoustic loss and the  Brillouin gain simultaneously, on the acoustic envelope. However; the impact of piezopotential envelope can not be observed directly.  
 By assuming that the variation of pump and Stokes along the waveguide length are much larger than $\alpha_{\rm new}^{-1}$ the acoustic envelope is 
 	\begin{equation}
	b\approx\frac{-i\Omega Q_{\rm b}}{\alpha_{\rm ac}\mathcal P_{\rm ac}+i\Omega \frac{\alpha_{\rm MEl}}{\alpha_{\rm ElM}}\widetilde Q_{\rm 2,piezo}}  A_{\rm p}A^{*}_{\rm s}.
\label{b_quantum}
			\end{equation}
 To observe the similarity between the above expression and the result obtained in \cite{otterstrom2023modulation}, we perform the following transformations together with the assumption that phase matching condition exists between the optical signals and the external acoustic wave
	\begin{equation}
	\begin{split}
 &\Gamma_{\rm m}= \alpha_{\rm ac} v_{\rm g},\\
 &\mathcal P_{\rm ac}=v_{\rm g} \mathcal E_{\rm ac},\\
& g_0^{*}=\frac{\Omega Q_{\rm b}}{\mathcal E_{\rm ac}},\\
&i\frac{1}{\mathcal E_{\rm ac}}\Omega \frac{\alpha_{\rm MEl}}{\alpha_{\rm ElM}}\widetilde Q_{\rm 2,piezo}= i\Delta \Omega-  G_{\rm AE},
  \end{split}
	\end{equation}
 where $\Gamma_{\rm m}$ is mechanical dissipation rate in $[\frac{1}{s}]$, $\mathcal E_{\rm ac}$ acoustic energy density per length and $g_0$ is an optomechanical coupling rate. The parameter of $G_{\rm AE}$ and $\Delta \Omega$ are related to real and imaginary parts of the piezoelectric impact respectively as shown in \cite{otterstrom2023modulation}. 
Then, the acoustic envelope take the following expression\\
	\begin{align}
b\approx\frac{-ig^{\ast}_0}{\Gamma_{\rm m}- G_{\rm AE}+i\Delta \Omega}  A_{\rm p}A^{\ast}_{\rm s},
	\end{align}\\
\noindent that is similar  to the result reported in \cite{otterstrom2023modulation}.

\end{document}